\documentclass[twocolumn,tighten]{aastex61}
\newcommand{\crit}{{\mathrm{c}}}
\newcommand{\sound}{{\mathrm{s}}}
\newcommand{\grav}{{\mathrm{g}}}
\newcommand{\turb}{{\mathrm{t}}}
\newcommand{\alfv}{{\mathrm{A}}}
\newcommand{\magb}{{\mathrm{b}}}

\newcommand{\Ms}{{\mathrm{M}_\odot}}
\newcommand{\Mspc}{{\mathrm{M}_\odot}~\textrm{pc}^{-1}}

\newcommand{\pc}{{\mathrm{pc}}}
\newcommand{\cc}{{\mathrm{cm^{-3}}}}

\received{December 14, 2016}
\accepted{August 29, 2017}
\shorttitle{Analytical CMF from filaments}
\shortauthors{Lee, Hennebelle \& Chabrier}
\begin{document} 
\title{Analytical core mass function (CMF) from filaments: \\
Under which circumstances can filament fragmentation reproduce the CMF?}

\correspondingauthor{Yueh-Ning Lee}
\email{yueh-ning.lee@cea.fr}

\author{Yueh-Ning Lee}
%\affil{Laboratoire AIM, Paris-Saclay, CEA/IRFU/SAp -- CNRS -- Universit\'e Paris Diderot, F-91191 Gif-sur-Yvette Cedex, France}
\affiliation{IRFU, CEA, Universit\'{e} Paris-Saclay, F-91191 Gif-sur-Yvette, France}
\affiliation{Universit\'{e} Paris Diderot, AIM, Sorbonne Paris Cit\'{e}, CEA, CNRS, F-91191 Gif-sur-Yvette, France}
\author{Patrick Hennebelle}
%\affiliation{Laboratoire AIM, Paris-Saclay, CEA/IRFU/SAp -- CNRS -- Universit\'e Paris Diderot, F-91191 Gif-sur-Yvette Cedex, France}
\affiliation{IRFU, CEA, Universit\'{e} Paris-Saclay, F-91191 Gif-sur-Yvette, France}
\affiliation{Universit\'{e} Paris Diderot, AIM, Sorbonne Paris Cit\'{e}, CEA, CNRS, F-91191 Gif-sur-Yvette, France}
\affiliation{Laboratoire de radioastronomie, UMR CNRS 8112, \'Ecole normale sup\'erieure et Observatoire de Paris, 24 rue Lhomond, F-75231 Paris cedex 05, France}

\author{Gilles Chabrier}
\affiliation{\'Ecole normale sup\'erieure de Lyon, CRAL, UMR CNRS 5574, Universit\'e de Lyon, F-69364 Lyon Cedex 07, France}
\affiliation{School of Physics, University of Exeter, Exeter EX4 4QL, UK}
 
\begin{abstract}
Observations suggest that star formation in filamentary molecular clouds occurs in a two-step process, with the formation of filaments preceding that of prestellar cores and stars. %Moreover, the observed core mass function (CMF) in filaments is similar to the standard Galactic initial mass function (IMF). 
%We study the fragmentation of filaments under various conditions by 
Here, we apply the gravo-turbulent fragmentation theory of \citet{HC08, HC09, HC13} to a filamentary environment, taking into account magnetic support. We discuss the induced geometrical effect on the cores, with a transition from 3D geometry at small scales to 1D at large ones. The model predicts the fragmentation behavior of a filament for a given mass per unit length (MpL) and level of magnetization. This CMF for individual filaments is then convolved with the distribution of filaments to obtain the final system CMF.
The model yields two major results: (i) the filamentary geometry naturally induces a hierarchical fragmentation process, first into groups of cores, separated by a length equal to a few filament Jeans lengths, i.e. a few times the filament width. These groups then fragment into individual cores. (ii) Non-magnetized filaments with high MpL are found to fragment excessively, at odd with observations. This is resolved by taking into account the magnetic field treated simply as additional pressure support). 
The present theory suggests two complementary modes of star formation: while small (spherical or filamentary) structures will collapse directly into prestellar cores, according to the standard Hennebelle-Chabrier theory, the large (filamentary) ones, the dominant population according to observations, will follow the afore-described two-step process.
\end{abstract}
%\keywords{      .         }
%________________________________________________________________________________

\section{Introduction}
Most prestellar cores are associated with filamentary structures in molecular clouds. 
A two-step scenario has been proposed \citep{Andre10} in which massive clumps inside molecular clouds first condense to form filaments, that then fragment to form cores. 
The Gould belt survey \citep{Konyves15} observed the Aquila and Polaris regions and found that, 
in the Aquila region where most filaments are supercritical, 
the prestellar cores are preferentially (75 \%) found in the filamentary structures. 
We refer to a filament as supercritical when gravity dominates over thermal energy. 
This is equivalent to a critical mass per unit length (MpL) $m_\lambda = 2c_\sound^2/G \approx 16 ~\mathrm{M}_\odot / \mathrm{pc}$ at 10 K \citep{Ostriker64}.
A core mass function (CMF) $dN/dM$ was inferred in this cloud with a high-mass end slope of -2.33 \citep{Konyves15}, compatible with the Salpeter value (-2.35).
This two-step picture is very plausible since anisotropy is easily amplified during gravitational collapse and a massive clump thus tends to collapse into a filament rather than into an individual core. 
This is also supported by the omnipresence of filaments in non-star-forming regions \citep[][Polaris]{Andre10}, 
implying that the formation of filaments probably precedes that of the stars \citep[][Musca]{Kainulainen16a}.  
Indeed magneto-hydrodynamic (MHD) simulations show that filaments can be formed even without gravity \citep{Hennebelle13, Federrath16}. 
%Is has also been suggested that magnetic field in filaments probably play an important role for the core formation. 
Many studies have also shown that filamentary structures are associated to the hierarchical condensation of matter and the formation of cores \citep[e.g.][]{Schneider79,Ballesteros99,Klessen00,Padoan01,Balsara01,Boldyrev02}. 
If this is a common mode of star formation, 
the important question to address is: 
does one single filament, with specific MpL, produce the CMF 
or is this latter produced by an ensemble of filaments with a given distribution of MpL ?
To address this question, 
we need to characterize the filament population 
and to understand the characteristics of filament fragmentation.

The initial mass function (IMF) is observed to be almost invariable in different environments under conditions typical of the Milky Way \cite[see reviews by][and references therein]{Chabrier03, Bastian10, Offner14}. 
Only under extreme atypical conditions has the IMF been predicted to differ from the universal Galactic one \citep{CHC14,Hopkins13a}.
Many theoretical efforts have been made to explain this observed universality of the IMF. 
Most of them are based on a fragmentation process that forms dense cores inside a turbulent medium by gravitational instability. 
It is then assumed that the final IMF is just a mapping of the CMF, 
as a result of a nearly mass independent process \citep[e.g.][]{Matzner00}. 
\citet{Padoan97} and \citet{Padoan02} first proposed that the universality of the IMF is a consequence of the statistical properties of the supersonically turbulent medium. 
\citet{HC08, HC09, HC13} used a different approach and applied the Press-Schechter formalism \citep{Press74} to non-linear fluctuations produced in a gravo-turbulent medium. 
In their formalism, the CMF results from the counting of self-gravitating structures at all scales in the medium. 
Using a qualitatively similar concept, \cite{Hopkins12a,Hopkins12b, Hopkins13a} applied the excursion set theory to explain the structure formation by consecutive instability barrier crossing from galactic scales down to star-forming scales. 

The ISM is known to be fractal with dimension $D \sim 2-3$ \citep{Elmegreen96,Elmegreen97,Elmegreen02,Sanchez05,Sanchez08,Kainulainen11,Gusev14,Lee16} from observations of large scale diffuse ISM or star-forming clumps. Simulations of turbulent ISM also yield similar results \citep{Federrath09,Padoan16,Konstandin16}, for which \citet{Fleck96} proposed theoretical explanations. The stellar population born within molecular clouds is expected to reflect the gas structure, including this fractal dimension. The fractal dimension of stars are derived from the local surface density, which exhibits a two-slope powerlaw, The slope at scales $\lesssim 0.1 ~\pc$ results from the close binary correlations and is smoothed out by stellar motions \citep{Larson95}, while the one at larger scales reflects the clustering of the stars and yields the fractal dimension. Observations of young stellar distributions.\citep{Gomez93,Larson95,Simon97,Nakajima98,Hartmann02,Enoch08,Stanke06,Hennekemper08,Kraus08,Sanchez09}, however, yield fractal dimensions $D \sim 1-2$, varying in different regions. \citet{Larson95} and \citet{Sanchez10} noted this apparent discrepancy between molecular gas and stellar distribution and discussed the possible different internal structure of molecular clouds or mechanisms that could happen during the star formation processes. These above studies show that there is variation of the fractal dimension among different star-forming environments, which differs from $D \sim 2-3$ and with $D\sim 1$ definitely not uncommon. Notably $D \sim 1$ in the Taurus region \citep{Larson95,Simon97,Hartmann02,Kraus08}, reflecting the filamentary geometry also observed with CO. 

Observations of star-forming regions with higher resolution indeed show that filaments are pronounced structures \citep{Arzoumanian11, Arzoumanian13,Konyves15,Koch15}, and are highly associated with core formation. 
Then, it is necessary to explore filament fragmentation to identify the impact of such geometry on core formation. In the present paper, we focus on this particular issue and seek to provide some physical insight on core formation when the mass is initially concentrated inside filamentary structures (molecular clouds) . We do not attempt to model the entire hierarchical structure formation from the diffuse medium to the cores. 
In the same vein, we focus in this study on the formation of the CMF and suggest direct implications on the final IMF based on the observed similarity between these two mass functions. This relation between the CMF and the final IMF is a subject of study on its own and remains a debated issue 
\citep[e.g][]{Offner14,Guszejnov15,Guszejnov16}.

Filament fragmentation has been previously studied in various contexts. 
\citet{Inutsuka92, Inutsuka97} studied the gravitational instability of perturbation growth of isothermal filaments in equilibrium at thermally critical MpL and found the most unstable mode to be about eight times the filament radius. 
\citet{Seifried15} showed with simulations that the fragmentation pattern depends on the turbulence and magnetic field morphology. 
\citet{Clarke16} useed SPH simulations and an analytical model to infer the link between growing perturbations and the mass accretion rate onto the filament. 
\citet{Gritschneder16} showed with AMR simulations that the geometrical oscillation allows the filament to fragment at any scale. 
In this paper, we apply the theory developed by \citet{HC08, HC09, HC13} to filament geometry and conditions. 
The Press-Schechter formalism was first applied to the star formation context by \citet{Inutsuka01} for studying the longitudinal fragmentation of thermally critical filaments. 
Our study aims at characterizing the magneto-gravo-turbulent fragmentation of filaments with different properties 
and at proposing possible explanations for the CMF observed in filamentary star-forming regions. 
We demonstrate with simple analytical models that the magnetic field probably plays an important role in regulating the filament fragmentation into star-forming cores. 
Note that this model only considers isolated filaments, and does not address intersections or hubs \citep[e.g.][]{Myers09} among filaments which are the observed sites of massive star formation. We will discuss this issue later. This is also consistent with the results of \citet{Sanchez07} of a fractal dimension $D \sim 2.8$ for OB star distribution, suggesting that massive stars are probably not associated with filamentary geometry and that their distribution might reflect that of larger scale clump or cloud structures.

\subsection{Outline of the model}
Since the model proposed in this work entails multiple steps, we first outline each step to guide the readers through the underlying ideas and also to identify the different assumptions. Generally speaking, the idea is to apply the scale-dependent gravo-turbulent fragmentation model proposed by Hennebelle \& Chabrier to filamentary star-forming environments, as suggested by observations. 

\subsubsection{Gravo-turbulent fragmentation}
The filamentary environment imposes constrains on the density, energy, and geometry for core formation. The shape of the core is constrained by the filament geometry, evolving from spheres at small scales to prolate ellipsoids at scales comparable to or larger than the filament width (\S\ref{st_filgeo}). The critical collapse density of a core is evaluated for all masses (sizes) according to the various scale-dependent supports (\S\ref{st_rhocrit}).  The mass distribution is then derived by counting unstable masses at all scales, given the density PDF within individual filaments (\S\ref{st_HC}). 

\subsubsection{Local conditions: the filament}
The model considers isolated filaments as initial conditions for the fragmentation, with a rather simple setup. The filament has a constant width of canonical value 0.1 pc (see Appendix \ref{appen_R} for relaxation on this condition), and constant mean density. Using a density profile with a central flat region and a powerlaw envelope, as suggested by observations, may introduce some corrections to the results of our model (Appendix \ref{appen_p}). 
Cylindrical virial equilibrium is applied, assuming that the filament is not in rapid radial contraction (\S\ref{st_filmod}). Aside from thermal and turbulent energy, the magnetic field is considered as the main pressure support; other processes such as cosmic rays, radiation and stellar jets can also possibly provide similar supporting pressure terms. Within the filament, the turbulence generates a lognormal density PDF that depends on the repartition of the different modes, yielding different results of the fragmentation (\S\ref{st_cases}). 

\subsubsection{Global conditions: the filament population}
The final, global CMF from a filamentary environment needs to be derived from a distribution of filaments (\S\ref{st_conv_mod}). The major parameters here are the filament mass per unit length (MpL )distribution and the level of magnetic intensity (or equivalent pressue support by any other mechanism). Motivated by observations, we explore several possibilities and present the results of the obtained global CMF (Section \ref{st_filpop}).\\

Having introduced the general ideas, the rest of the paper is organized as follows: 
The model of filament fragmentation is introduced in Section 2. 
Readers who do not wish to go into details of the formalism are invited to skip to Section 3, 
where we discuss the fragmentation as results of different filament conditions. 
Convolution with the filament population is done in Section 4. 
Finally, we conclude in Section 5.

%________________________________________________________________________________

\section{CMF inside individual filaments: the formalism}

The Hennebelle \& Chabrier formalism \citep{HC08, HC09,HC13} is a statistical approach that counts the self-gravitating structures at all scales $l$ within a gravo-turbulent medium, 
while taking into account the scale dependence of the supporting thermal, turbulent and magnetic energy. 
The main quantities entering the formalism are:
\begin{itemize}
\item $P(\delta)$: The probability density function (PDF) of the log overdensity, $\delta = \log(\rho/\rho_0)$, 
where $\rho$ and $\rho_0$ are the local and mean density of the filament, respectively. 
Note that this is the local density distribution within an individual filament. 
\item $\sigma^2(l)$: The Fourier transform of power spectrum of density fluctuations of size $l$, where $l$ is the relevant scale of the density perturbation, inside the filament. 
This determines the width of the log density PDF smoothed at scale $l$. 
\item $\rho_\mathrm{c}(l)$: The critical density of collapse as function of the perturbation length $l$ (or corresponding mass $M_l^\crit$),
that depends on the geometry and the type of support considered.
\end{itemize}

\startlongtable
\begin{deluxetable*}{Ll}
\tablecaption{Variables \label{table_vars}}
%\tablehead{
%\colhead{Variable} & \colhead{Description}
%}
\tablehead{
\mathrm{Variable} & Description
}
%\colnumbers
\startdata
%\begin{table*}[ht!]
%\caption{Variables}
%\label{table_vars}
%\centering
%\begin{tabular}{l l}
%\hline\hline
%Variable & Description \vspace{.5mm}\\
%\hline 
\multicolumn2c{\bf Constants} \\
G                 & Gravitational constant \\
c_0              & Isothermal sound speed 200 m s$^{-1}$ at 10 K \\
m_\crit         & Thermally critical MpL $2c_0^2/G = 16~ \Mspc$ at 10 K\\  
\rho^\mathrm{crit}& Critical density of polytropic equation of state transition $2.5 \times 10^5 ~\cc$ \\
\gamma_1/\gamma_2 & Polytropic index of gas equation of state below/above the critical density:  0.7/1.1\\    
\eta_\turb     & Kolmogorov exponent of the velocity dispersion vs. scale relation \\
\gamma_\magb  &Exponent of the magnetic field vs. density relation \\
m                 & smoothing parameter of the eos transition\\
n                  & smoothing parameter of the clump geometrical transition \\
\hline
\multicolumn2c{\bf Filamentary cloud parameters} \\
\mathcal{N}_\mathrm{sys} & System CMF, number of cores per unit mass \\
\mathcal{N}_\mathrm{c}^{m_\lambda} & CMF of filament with MpL $m_\lambda$, number of cores per unit length per unit mass  \\
\mathcal{N}_\mathrm{f} & Filament mass distribution, number per unit MpL, spatial number density not considered\\
m_\lambda  & Filament average MpL \\
R                  & Radius of the filament (half width) \\
L_\mathrm{tot}  & Total length of the filament \\
V_\mathrm{tot}  & Total volume of the filament $\pi R^2 L_\mathrm{tot}$\\
M_\mathrm{fil} & Total mass of filament $m_\lambda L_\mathrm{tot}$ \\
\rho_0          & Filament average density $m_\lambda/(\pi R^2)$ \\
M_0              & Characteristic mass of the filament, chosen to be $4 m_\lambda R / 3$ \\
\sigma_0     & Standard deviation of the lognormal density PDF at small scales ($l\rightarrow 0$) \\
v_\turb^\ast  & One-dimensional turbulence dispersion at scale $R$\\
\mathcal{M}_\mathrm{1D}  & One-dimensional sonic Mach number $v_\turb^\ast/c_0$ at scale $R$  \\
B_0               & Magnetic field at filament average density\\
v_\alfv^\ast          & Alfv\'en velocity at filament average density $B_0/ \sqrt{4\pi\rho_0}$ \\
V_\alfv          &  Alfv\'en velocity at filament average density normalized by sound speed $v_\alfv^\ast/c_0$ \\
\mathcal{M}_\alfv  & Alfv\'enic Mach number at filament average density $\sqrt{3}v_\turb^\ast/v_\alfv^\ast = \sqrt{3}\mathcal{M}_\mathrm{1D}/V_\alfv$ \\
K_\mathrm{crit} & Constant of eos transition at filament density \\
\hline
\multicolumn2c{\bf Local variables} \\
\rho              & Density \\
\delta           & Logarithmic density contrast $\log(\rho/\rho_0)$ \\
\rho_\crit (l)     & Density of the virially critical core, as function of scale \\
\sigma (l)         & Standard deviation (half width) of the lognormal density PDF, as a function of the smoothing scale \\
l                   & Semi-major axis of the prolate clump oriented along the filament \\
r                   & Semi-minor axis  of the prolate clump oriented along the filament \\
\eta               & Aspect ratio of the prolate clump $l/r$\\
M_l^\crit     & Mass of the virially critical clump at scale $l$ \\
c_\sound ($\rho$)     & Sound speed, function of local density \\
v_\turb (l)         & One-dimensional turbulent velocity dispersion, function of scale \\
v_\alfv  ($\rho$)          & Alfv\'en velocity $B(\rho)/\sqrt{4\pi\rho}$, function of local density \\
E_\grav        & Gravitational potential energy of the clump\\
w_\grav(\eta)        & Geometrical factor of the gravitational potential energy, function of $\eta$ \\
\hline
\multicolumn2c{\bf Normalized variables} \\
T                  & Temperature $c_\sound^2/c_0^2$ \\
\widetilde{\rho}     & Density $\rho/\rho_0$ \\
\widetilde{M}       & Critical clump mass $M_l^\crit/M_0$ \\
\widetilde{l}         & Semi-major axis $l/R$ \\
\widetilde{r}         & Semi-minor axis $r/R$ \\
\Lambda      & MpL $8m_\lambda/15m_\crit$ \\
\enddata
%\tablecomments{}
\end{deluxetable*}

%\hline
%\hline
%\end{tabular}
%\end{table*}

\subsection{The density PDF}
Since we are discussing fragmentation in supercritical, turbulent filaments, we use
a lognormal PDF, knowing that in the limit of small density fluctuations, $\delta \rho \ll \rho_0$, 
a lognormal PDF approaches a gaussian one:
\begin{equation}
P_l(\delta) = {1 \over \sqrt{2 \pi \sigma^2(l)}} \exp{\left( {-\left(\delta+ \sigma^2(l)/2\right)^2 \over 2\sigma^2(l)}\right)}.
\end{equation}
The function $P_l(\delta)$ describes the probability of finding an overdensity $\delta$ at smoothing scale $l$. 
A lognormal density PDF is obtained analytically and numerically for a nearly isothermal turbulent medium \citep[e.g.][]{Vazquez94,Padoan97}\footnote{Although we consider lognormal distribution within the filament, in Section \ref{st_fila_pop}  we sum up over a distribution of filaments, which leads to a powerlaw distribution of the density PDF as indeed suggested by observations \citep[see Figure 15 of][]{Konyves15}.}. Deviations from lognormality occur for non-isothermal gas equation of state (eos) \citep[e.g.][]{Scalo98,Passot98,Federrath15} or large Mach number \citep{Hopkins13b}, but in this paper we restrict our discussion to the simplest PDF expression which is sufficient to illustrate the fragmentation behavior in dense filamentary molecular clouds. Note, however, that any form of density PDF can be used in our formalism if necessary but analyticity and simplicity of the calculations will be lost.

The PDF dispersion width decreases with increasing smoothing scale, such that
\begin{equation}
\label{eq_sig}
\sigma^2(l) = \sigma_0^2 \left(1-\left({2l\over L_\mathrm{tot}}\right)^{2\eta_\mathrm{t}} \right),
\end{equation}
where $L_\mathrm{tot}$ is the total length of the filament. 
This scaling relation is obtained by integrating the density power spectrum, filtered by a window function, from the wave number corresponding to the integral scale $k=2\pi/L_\mathrm{tot}$ to infinity \citep{Stutzki98, Bensch01,Hennebelle07,HC08}. 
The dispersion tends to zero at the scale of the whole filament, 
reflecting the fact that density fluctuations vanish at this level,
and approaches $ \sigma_0^2$ at small scales. 
For the turbulent velocity power spectrum index, we take the Kolmogorov value $\eta_\turb =  0.45$.
This value is also consistent with the observed density power spectrum of a supercritical filament, characterized by a logarithmic powerlaw slope $\sim -1.9 = - (1+2\eta_\mathrm{t})$ \citep{Roy15}. 
Numerical simulations suggest that the variance of the PDF is related to the 1-D turbulent Mach number as
\begin{equation}
\label{eq_sig0}
\sigma_0^2 = \log{(1+3b^2\mathcal{M}_\mathrm{1D}^2)}, 
\end{equation}
where $b\simeq 0.5$ corresponds to equipartition between the various modes of turbulence \citep{Federrath10}. 
As a natural consequence of turbulence, 
the higher the Mach number, the wider the PDF. 
The variables used in our model are listed in Table \ref{table_vars}. 

%This is a reasonable assumption, since observed filaments are well described by a Plummer profile, with most of the mass concentrated in the central flat region. 

Most observed filaments are well described by a density profile given by
\begin{equation}\label{eq_plummer}
\rho(r) = {\rho_0 \over \left[1+\left({r \over r_0}\right)^2\right]^{p/2}},
\end{equation}
that reduces to the static Ostriker profile with $p=4$, while $p \lesssim 2$ is often reported by observations. 
For sake of simplicity, we only consider a constant radial density profile in the filament. 
Taking into account a more realistic profile will yield corrections of order unity (see Appendice \ref{appen_n} and \ref{appen_p} for discussions), comparable to other uncertainties in the model (e.g. geometrical factors in the virial condition).

\subsection{Filament geometry}\label{st_filgeo}
The core-forming density fluctuations forming inside the filament are geometrically limited by the material that is available. 
We consider ellipsoidal fluctuations, with the relevant perturbation scale $l$ being the semi-axis along the filament {\it longitudinal} direction (see Appendix \ref{appen_p} for discussions on spherical fluctuations inside a filament with a powerlaw envelope). 
The relevant scale $r$ along the filament {\it radial} direction scales as $l$ at small scales and approaches the filament radius $R$ at large $l$. 
In order to describe this geometrical evolution we use the form 
\begin{equation}
r = \left( l^{-n} + R^{-n} \right)^{-{1\over n}},
\label{eq_r_of_l}
\end{equation}
where $n$ is a parameter governing the smoothness of the transition.
We take $n=2$ in the following calculations, 
and the dependence of the results on this parameter is discussed in Appendix \ref{appen_n}. 
This relation naturally yields spherical fluctuations at small scales since they are not sensitive to the filamentary geometry, 
and prolate ellipsoidal fluctuations at scales comparable with the filament width. 
This geometrical transition also implies a transition from 3D to 1D in the PDF. Accordingly,  
we modify the density PDF width such that
\begin{equation}
\sigma^2(l) = \sigma_0^2 \left(1-\left({2l\over L_\mathrm{tot}}\right)^{2\eta_\mathrm{t}} \right)\left({2r+l \over 3l}\right).
\label{eq_dPDF}
\end{equation}
Given our limited understanding of the anisotropic density PDF, 
we consider that such a simplifying formula is reasonable. 
At small scales, $r \sim l$, the density PDF remains unmodified. 
At large scales, $r \sim R \ll l$, this equation yields a width for the PDF approximately 1/3 of the original width in the case of a very elongated clump, 
reflecting the fact that the scale varies in only one of the three dimensions.

\subsection{Density threshold to formation of an unstable core}\label{st_rhocrit}
To determine the critical density for a collapsing core, we use the virial theorem to compare the gravitational energy vs the thermal, turbulent, and magnetic energies of the density fluctuations prone to gravitational instability (which are the core progenitors). 
We integrate in a given volume the inner product of the momentum equation and the position vector \citep[e.g.][]{Collins78}. 
Equilibrium is found by setting the time dependent terms to zero. Neglecting the surface terms, one gets
\begin{equation}
-E_\grav(M_l^\crit) = 3 M_l^\crit \left(  c_\sound^2(\rho)  + v_\turb^2(l) + {v_\alfv^2(\rho)\over 6} \right),
\label{eq_vir}
\end{equation}
where $E_\grav$ is the gravitational energy, 
$M_l^\crit$ is the critical mass at scale $l$, 
$c_\sound (\rho) $ is the local sound speed, calculated with a barotropic eos (see below),
$v_\turb$ the scale-dependent 1-D turbulent velocity, 
and $v_\alfv$ is the density-dependent Alfv\'{e}n velocity.
Although other sources of energy, such as cosmic rays or thermal radiation, might also provide some support, they have not been observed to be correlated with filaments, in contrast to magnetic fields, with magnetic energy broadly comparable to turbulent energy. 

Assuming a uniform density, 
the gravitational energy of an ellipsoid takes the form \citep{Parker54, Neutsch79}
\begin{equation}\label{egrav}
-E_\grav(M_l) = {3 \over 5} {GM_l^2 \over l} w_\grav(\eta),
\end{equation}
where $w_\grav$, the geometrical factor depending on the core aspect ratio $\eta = l/r$, 
has the form
\begin{equation}
w_\grav(\eta) = {\eta\log{(\sqrt{\eta^2-1}+\eta)} \over \sqrt{\eta^2 -1}} = {\eta\sinh^{-1}(\sqrt{\eta^2 -1}) \over \sqrt{\eta^2 -1}} 
\end{equation}
for prolate ellipsoids with $\eta>1$.
More details on the function $w_\grav$ are discussed in Appendix \ref{appen_w}.

The turbulent velocity depends on the scale as 
\begin{equation}
v_\turb(l) = v_\turb^\ast \left( {l \over R}\right)^{\eta_\turb},
\label{eq_vturb}
\end{equation}
where $v_\turb^\ast$ is the 1-D turbulent velocity at the filament scale, 
since $R$ is assumed to be the injection scale of the turbulent energy. 
We assume that the turbulent velocity only depends on $l$ rather than fluctuations in the three directions to simplify the problem, since the longitudinal scale $l$ for fluctuations is the most relevant one. 
As mentioned earlier, the value of $\eta_\turb$ is typically 0.3-0.5 and we take 0.45 throughout this study, 
consistent with the observed density power spectrum index \citep{Roy15}. 
The magnetic field scales as 
\begin{equation}
B = B_0\left({\rho \over \rho_0}\right)^{\gamma_\magb}, 
\end{equation}
where $B_0$ is the magnetic field at the filament mean density $\rho_0$ and the exponent $\gamma_b$ has a typical value $<$2/3. 
This upper limit corresponding to flux freezing during isotropic contraction \citep{Crutcher10,LiP15}. 
The Alfv\'{e}n velocity, $v_\mathrm{A}^2 \propto B^2 / \rho$, thus scales with density as 
\begin{equation}
v_\alfv^2  = v_\alfv^{\ast 2} \left({\rho \over \rho_0}\right)^{2\gamma_\magb-1}.\label{eq_vmag}
\end{equation}
We take $\gamma_b=0.5$ for our fiducial value.  
This is a reasonable choice in the case of anisotropic contractions along magnetic field lines when the mass-to-flux ratio is not significantly larger than its critical value. 
The  data presented in the Figure 2 of \citet{Li15} gives a value $\gamma_b \simeq 0.5$ for pure gravo-magnetic simulations with mass-to-flux ratio a few times unity. 
The above simple relation thus provides the required magnetic field density dependence deriving from flux conservation in a turbulent medium. 
In this study, full MHD dynamics is not considered and the magnetic field is treated simply as a pressure term. 
For example the topology of the magnetic field  which may have an influence on the stability condition \citep{Fiege00a, Fiege00b} is not
considered. 
Magnetic fields outside supercritical filaments are usually observed to be perpendicular to the filament axis. 
Although not readily observed, the field inside filaments should follow similar directions or be randomly distributed, depending on the relative strength of turbulence.

Given the above equations, the critical mass $M_l^\crit$ and the corresponding critical density $\rho_l^\crit$ can be determined at any given scale $l$.
Normalizing the various quantities as $T = c_\sound^2/c_0^2$, 
$\mathcal{M}_\mathrm{1D} = v_\turb^\ast / c_0$, the 1D turbulent Mach number at the scale of the filament width,
$V_\alfv = v_\alfv^\ast / c_0$, the Alfv\'en velocity normalized by the sound speed $c_0$ for the filament mean density, 
$\widetilde{l} = l/R$, 
$\widetilde{\rho} = \rho \pi R^2 / m_\lambda$, 
and $\widetilde{M} = 3M_l^\crit / (4m_\lambda R) = M / M_0  = \widetilde{\rho}  \widetilde{r}^2  \widetilde{l} $, 
we obtain the dimensionless equation for virial equilibrium:
\begin{equation}
\widetilde{M} = {\Lambda^{-1}\widetilde{l} \over w_\grav(\eta)} \left( T(\widetilde{\rho}) + \mathcal{M}_\mathrm{1D}^2 \widetilde{l}^{2 \eta_\turb} +{V_\alfv^2\over 6} \widetilde{\rho}^{2\gamma_\magb -1} \right),
\label{eq_M}
\end{equation}
where $\Lambda = 4 Gm_\lambda/ (15 c_0^2) = 8 m_\lambda /(15 m_\crit)$ is a constant of a given filament. 
The critical value $m_\crit$ is the maximal MpL that can be reached with purely thermal support. 

The thermodynamic of the gas is described by the same barotropic eos as in \citet{HC09}
\begin{equation}
T(\widetilde{\rho}) = \left[ \widetilde{\rho}^{(\gamma_1-1)m} + K_\mathrm{crit}^m~ \widetilde{\rho}^{(\gamma_2-1)m} \right]^{1/m}, \label{eq_eos}
\end{equation}
with $\gamma_1 \simeq 0.7$ and $\gamma_2 \simeq 1.1$ with a transition at the critical density $\rho^\mathrm{crit} \simeq 10^{-18} \mathrm{g}~ \mathrm{cm}^{-3}$, i.e., 
$n^\mathrm{crit} \simeq 2.5 \times 10^5 ~\mathrm{cm}^{-3}$ \citep{Larson85}. 
In the following, we take $m = 3$. 
The constant
\begin{equation}
K_\mathrm{crit} =  \left({\rho^\mathrm{crit} \over \rho_0} \right)^{\gamma_1-\gamma_2}
\end{equation}
follows from the definitions.

\subsection{The Hennebelle \& Chabrier formalism}\label{st_HC}
The basic idea of this theory relies on the counting of all the unstable mass in the medium at all scales. 
A gravitationally unstable region is counted only when it does not contain any other smaller unstable mass, and thus the model properly accounts for the known ``cloud-in-cloud" problem in the Press-Schechter formalism \citep[see][]{HC08}. 
This corresponds to the last crossing in the excursion set theory \citep{Hopkins12a,Hopkins12b,Hopkins13a}. 
We stress that, although turbulence velocity enters the virial equilibrium condition, turbulence in the Hennebelle-Chabrier formalism  should not be regarded as a static support but rather as a dynamical process \citep{CH11}. It acts instantaneously against the gravity by sweeping out density fluctuations which would otherwise have collapsed, enabling us to \emph{statistically} determine the number of genuinely unstable fluctuations in a \emph{ turbulent medium}. Fluctuations for which gravity eventually dominates will collapse on their own and decouple from the environment, leading eventually to prestellar cores. At this stage, turbulence will have since long been dissipated. 
%The radiative cooling processes allow energy to be radiated away and magnetic flux is lost through ambipolar diffusion. 
We do not model the details of the core collapse phase itself which will lead to the final IMF. 

First, all regions with density higher than $\rho_l^\crit$ at scale $l$ are virially unstable and should collapse into cores, 
yielding the total collapsing mass
\begin{eqnarray}
\label{eq_int_dens}
M_\mathrm{tot}(l) &=&  V_\mathrm{tot} \int\limits_{\delta_c}^\infty \rho_0 \exp{(\delta)} P(\delta) d \delta \\
&=& L_\mathrm{tot} \int\limits_{\delta_\crit}^\infty m_\lambda \exp{(\delta)} P(\delta) d \delta, \nonumber
\label{eq_tot_mass}
\end{eqnarray}
where $V_\mathrm{tot} = \pi R^2 L_\mathrm{tot}$ is the total volume of the filament. 
On the other hand, at scale $l$,
there are still smaller fluctuations that can reach higher densities than the critical value.
Therefore, cores can form at masses smaller than the critical mass $M_l^\crit$.
The total mass at scale $l$ included in collapsing structures from zero to the critical mass thus corresponds to:
\begin{equation}
M_\mathrm{tot}(l) =  L_\mathrm{tot} \int\limits_0^{M_l^\crit} M^\prime  \mathcal{N}_\crit^{m_\mathrm{\lambda}}(M^\prime) P(M^\prime |M_l^\crit) dM^\prime,
\label{eq_int_mass}
\end{equation}
where $\mathcal{N}_\crit^{m_\mathrm{\lambda}}$ denotes the prestellar CMF in a filament of MpL $m_\mathrm{\lambda}$. 
As discussed in \citep{HC08}, for the sake of simplicity, we suppose that the probability that a core of mass $M^\prime$ is located inside $M_l^\crit$ is $P(M^\prime |M_l^\crit)= 1$, 
meaning that an unstable clump is always contained in a larger unstable clump.

Equating equations (\ref{eq_int_dens}) and (\ref{eq_int_mass}) and taking the derivative with respect to $M_l^\crit$ yields
\begin{eqnarray}
\label{eq_MF}
\mathcal{N}_\crit^{m_\mathrm{\lambda}}(\widetilde{M}) 
&=& {-m_\lambda \over M_0^2\widetilde{M}} {d\widetilde{l}\over d\widetilde{M}}\\
&& \left( \exp{(\delta_\crit)}P_l(\delta_c){d\delta_\crit \over d\widetilde{l}}- \int\limits_{\delta_\crit}^\infty \exp{(\delta)} {dP_l(\delta)\over d\widetilde{l}} d \delta \right) \nonumber
\end{eqnarray}
in the normalized form. 

Taking into account now the time-dependence of the collapse condition as in \citet{HC13}, 
we multiply the right-hand side of equation (\ref{eq_MF}) by a factor 
\begin{equation}
\label{eq_time_dep}
{1\over \phi_t} \sqrt{\widetilde{\rho}} \nonumber, 
\end{equation}
where $\phi_t \simeq 3$ is a constant of normalization. 
The physical meaning of this term is that density fluctuations are constantly replenished at all scales and are stationary, 
while cores of different densities collapse at time scales proportional to their respective free-fall time. 
As a result, 
small-scale cores form at a higher rate than the large ones, yielding an extra weighting in the mass function \citep[see][]{HC13}. 
As mentioned earlier, we do not address the collapse of the cores, which is explored in other studies \citep{Guszejnov15, Guszejnov16}. Collapsed cores are decoupled from the filament material and have no strong impact on the subsequence formation of other cores. This is supported by Figure 15 of \citet{Konyves15}, where the column density PDF shows that the mass inside collapsed cores is small compared to that within the filaments.

Because of the dependence of fluctuations upon the ellipsoidal geometry and the density-dependence of the thermal and magnetic energies, 
this mass function can not be written as an explicit function of the mass $\widetilde{M}$. 
By taking the derivative of equation (\ref{eq_M}) with respect to $\widetilde{l}$, 
we get the expressions
\begin{equation}
{d \widetilde{M} \over d \widetilde{l}}  =  {B\over C},
\label{eq_dBdC}
\end{equation}
where 
\begin{eqnarray}
\label{eq_BCD}
B &=& {\widetilde{M} \over \widetilde{l}}(w_\grav(\eta) - w_\grav^\prime(\eta)\eta\widetilde{r}^n) + \Lambda^{-1}\mathcal{M}_\mathrm{1D}^2 2\eta_\turb\widetilde{l}^{2\eta_\mathrm{t}} \\
&&- \Lambda^{-1}{dD\over d\widetilde{\rho}}\widetilde{\rho}(1+2\eta^{-n}), \nonumber\\
C &=& w_g(\eta) - {\Lambda^{-1} \over \widetilde{r}^2}{dD\over d\widetilde{\rho}}, \\
D &=& \left(\widetilde{\rho}^{(\gamma_1-1)m} + K_\mathrm{crit}^m\widetilde{\rho}^{(\gamma_2-1)m}\right)^{1\over m} +{V_\alfv^2\over 6}\widetilde{\rho}^{2\gamma_\magb-1}, 
\end{eqnarray}
and
\begin{equation}
\label{eq_dddl}
{d \delta \over d \widetilde{l}} = {d \log{\widetilde{\rho}} \over d \widetilde{l}} 
= {1 \over \widetilde{M}} {d \widetilde{M} \over d \widetilde{l}} - {1\over \widetilde{l}}(1+2\eta^{-n}).
\end{equation}

Notice that if we take the limit $n \rightarrow \infty$, 
the low-mass part of the CMF naturally approaches the spherical result of \citet{HC13} with $r \rightarrow l$, 
while the high-mass end is a purely elliptical CMF with semi-minor axis $r = R$ for all cores. 
The HC CMF for spherical cores is a power law at large scales and almost a lognormal distribution around the peak mass, determined by the mean density, the level of thermal and magnetic support, and the Mach number. 
In contrast, in the filament configuration, elongated structures become very unstable with increasing scales and thus more prone to fragmentation. 
Fragments with very large masses are therefore difficult to form under such a geometry. 
This is a natural consequence of the mass concentration in the 1D filament. 
This result still holds even when the density powerlaw envelope of the filament is considered (see Appendix \ref{appen_p}).

%\subsection{Results}
%________________________________________________________________________________

\section{CMF inside individual filaments: results}
We now present the results. First, we describe the filament models that are applied as the core-forming environment. Secondly, we discuss the properties of the resulting CMF.
\subsection{Filament model}\label{st_filmod}
Filaments exhibit a wide range of properties. 
The results of \citet{Arzoumanian11} show that filaments, while all having a width around 0.1 pc, span about three orders of magnitude in column density. 
Several studies have also shown that filaments are magnetized \citep{Li15,Zhang15,Pillai15}, 
while no widely accepted conclusion has been reached on the intensity of the magnetic field. 
Here, we examine various cases illustrating different conditions inside a filament to illustrate how the filament CMF depends on the environment. 
We consider only supercritical filaments, that have supersonic total velocity dispersion (thermal plus turbulent), 
since only these filaments are prone to gravitational instability and undergo fragmentation \citep[e.g.][]{Inutsuka92, Inutsuka01, Konyves15}. 
Some examples of CMF inside filaments with different MpL and magnetization levels are displayed in the upper panel of Figure \ref{fig_cases}. 
We use the canonical length 10 pc and diameter 0.1 pc for the filaments. 
As will be shown in the results, fragmentation occurs essentially at scales smaller than the filament width. This means that the filament length $L_\mathrm{tot}$, entering only through equation (\ref{eq_sig}), has no strong impact on the results as long as the filament aspect ratio is sufficiently large.
The non-thermal energies are calculated using the cylindrical virial equilibrium \citep{HA13} of the filament and energy partition
\begin{eqnarray}
\label{eq_vir_cyn}
{GM_\mathrm{fil}^2 \over L_\mathrm{tot}} &=& 2M_\mathrm{fil}(c_\mathrm{s}^2 + {v^\ast_\turb}^2+{v^\ast_\alfv}^2/6). %\\
%{E_\mathrm{mag} \over E_\mathrm{nt}} &=& {{v^\ast_\alfv}^2 \over {v^\ast_\alfv}^2 + 3{v^\ast_\turb}^2},
\end{eqnarray}
%where $E_\mathrm{mag}$ and $E_\mathrm{nt}$ and the magnetic and total non-thermal energies, respectively. 
We calculated the energy balance in 2D and use this result for the 3D model, 
assuming isotropy of the pressures.  
Given the polytropic eos (equation (\ref{eq_eos})) and $0.1 ~\pc$ width, the filament supported purely by thermal energy has critical MpL $30 ~\Mspc$. This equilibrium condition for a quasi-stationary accreting filament is supported by the fact that filaments are stable structures compared to the occurance of density fluctuations and core formation, and surface terms are not dominating \citep[see e.g.][]{HA13}. Moreover, the turbulent velocity is observed to roughly satisfy equation (\ref{eq_vir_cyn}) \citep{Arzoumanian11}.
The condition remains valid during the formation of cores since, as mentioned earlier, the mass fraction within the cores is small compared to the mass in the filament \citep[see Figure 15 of][]{Konyves15}. 
We emphasize that equation (\ref{eq_vir_cyn}) is applied to the filament and assumes {\it global} mechanical equilibrium, 
while equation (\ref{eq_vir}) concerns the density fluctuations within the filament. 
As mentioned earlier, we assume a uniform density profile within filaments, which is a reasonable approximation since observations show that filaments have a dense central cusp and a steeply decreasing envelope \citep[$\rho(r) \sim r^{-2}$][]{Arzoumanian11}. Furthermore, the smoothing transition from spherical to ellipsoidal profiles mimics to some extent the effect of a powerlaw envelope, and the maximal fragment changes only slightly with different smoothing functions (Appendix \ref{appen_n}).   Considering the mass contribution from the filament envelope only results in a small correction (Appendix \ref{appen_p}). 
The Alfv\'enic Mach number 
\begin{equation}
\mathcal{M}_\alfv = \sqrt{3}\,v^\ast_\turb/v^\ast_\alfv% = \sqrt{{E_\mathrm{nt} \over E_\mathrm{mag}} -1}
\label{eq_Ma}
\end{equation}
can be calculated accordingly. 
We have carried out calculations for $\mathcal{M}_\alfv =1$ and $2$, 
close to the typical values found in \citet{Crutcher10}. 
For a filament predominately supported by turbulence, equation ({\ref{eq_vir_cyn}) leads to
\begin{equation}
Gm_\lambda \approx {2\over 3} \mathcal{M}_\alfv^2 {v^\ast_\alfv}^2  = {2\over 3} \mathcal{M}_\alfv^2 {B_0^2 \over 4\pi\rho_0}.
\end{equation}
With $m_\lambda = \pi R^2 \rho_0$, we obtain
\begin{eqnarray}
B_0 &\approx& \sqrt{6G} {m_\lambda \over\mathcal{M}_\alfv  R}  \\
&\approx& 22 ~\mu G~ \mathcal{M}_\alfv^{-1} \left({m_\lambda \over 16 ~\Mspc}\right)\left({R \over 0.1 \pc} \right)^{-1}. \nonumber
\end{eqnarray}
The density of a filament follows 
\begin{equation}
n = {m_\lambda \over \pi R^2m_\mathrm{p}} = 10^4~ \cc \left({m_\lambda \over 16 ~\Mspc}\right)\left({R \over 0.1 \pc} \right)^{-2},
\end{equation}
where $m_\mathrm{p}$ is the average molecular weight. 
Given these relations, the magnetic field density dependence in our filament model reads 
\begin{equation}
B_0 \approx 22~\mu G ~\left({n \over 10^4~\cc}\right)\mathcal{M}_\alfv^{-1} \left({R \over 0.1 \pc} \right).
\end{equation}
This relation is not far from the dependence observed by \citet{Crutcher10} and verified numerically by \citet{LiP15} for dense clumps:
\begin{equation}
B = 42~\mu G ~\left({n \over 10^4~\cc}\right)^{0.65},
\end{equation}
implying that the range of magnetic field strengths that we consider is reasonable. 
The difference in density dependence can be easily compensated by a slight increase of $\mathcal{M}_\alfv$ with the MpL. 
It should also be kept in mind that the observed relation has itself a large dispersion. 
Moreover, note that here we are discussing average quantities in filaments, which have internal fluctuations that eventually form cores, 
while the above relation is inferred from dense clumps.

\subsection{CMF variation with filament properties}\label{st_cases}
\begin{figure*}[]
\centering
\plottwo{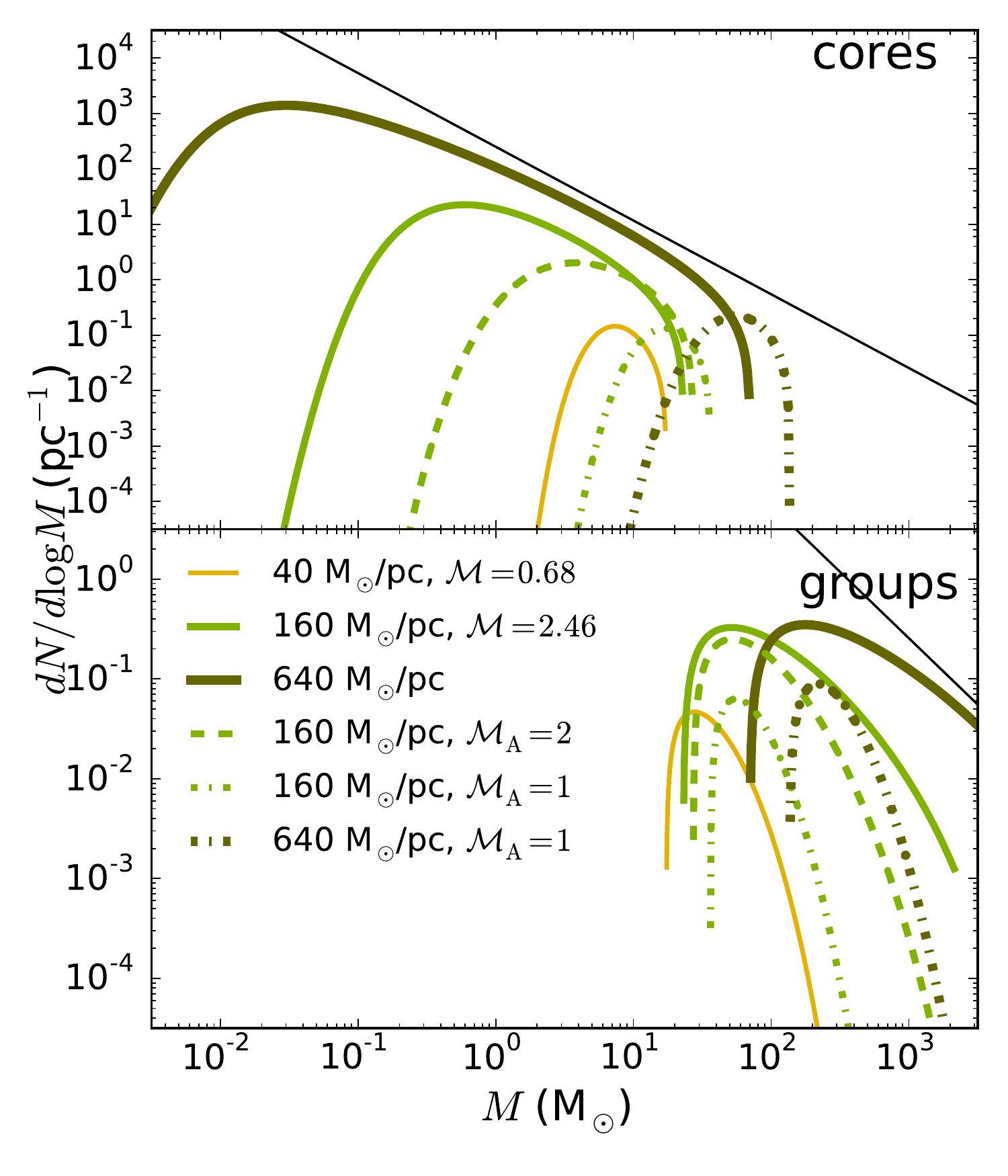}{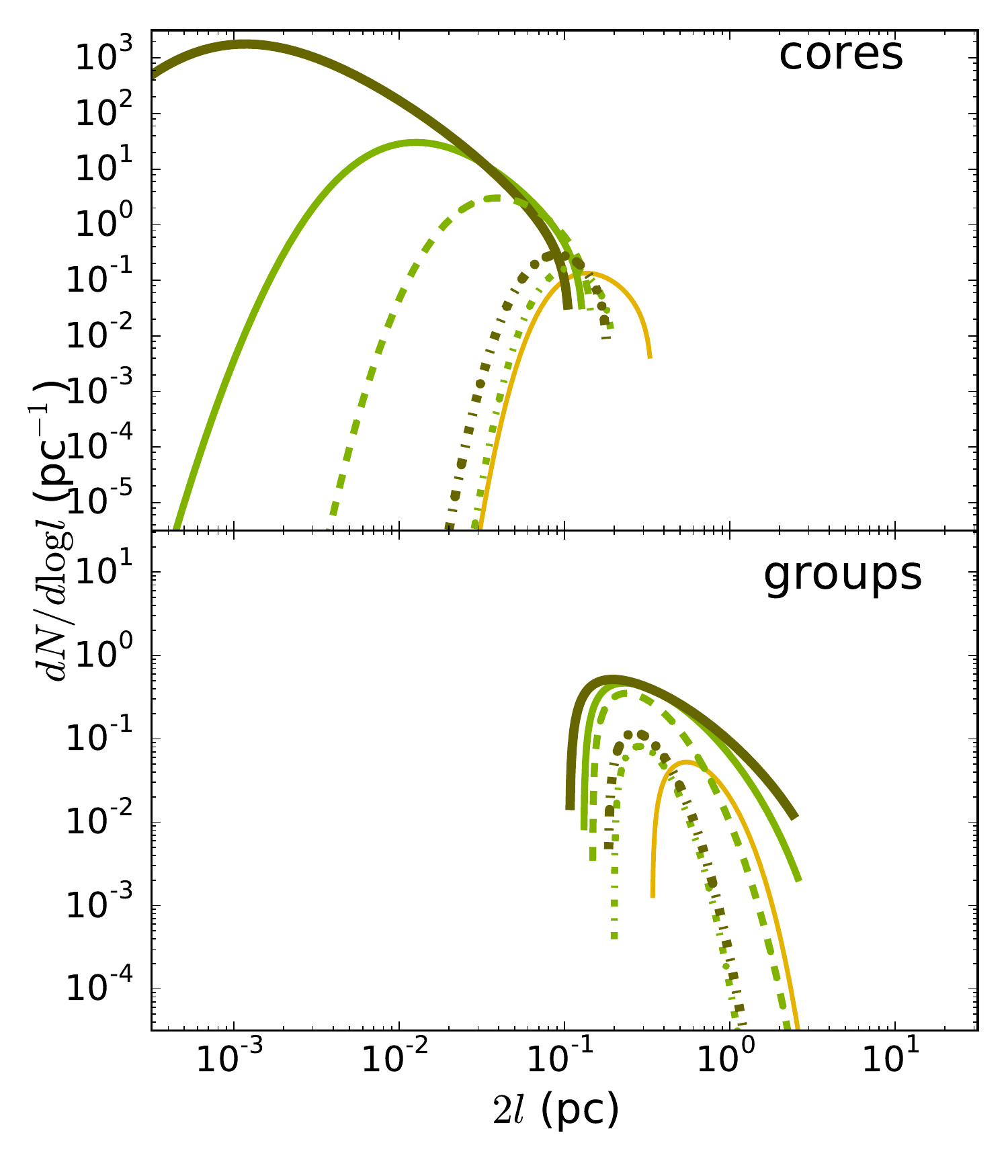}
\caption{\emph{Left:} CMF inside filaments of different MpL and magnetization. Solid curves show CMF in non-magnetized filaments of 40 (yellow), 160 (light green), and 640 (dark green) $\Mspc$ with increasingly wide curves. Dashed and dot-dashed light green curves show CMF in 160 $\Mspc$ filament with $\mathcal{M}_\alfv = 2$ and $1$. Dot-dashed dark green curve shows CMF in 640 $\Mspc$ filament with $\mathcal{M}_\alfv = 1$. The straight line shows the slope -1.33 measured by \citet{Konyves15} in Aquila. We use canonical values of 10 pc length and 0.1 pc diameter for all filaments.
\emph{Right:} Number distribution $dN/d\log{l}$ of self-gravitating fluctuations within filaments, just at the edge of collapse, as a function of their characteristic major axis $2l$. Same cases as in the left panel. These sizes of fluctuations just before collapse can be regarded as the typical wavelengths of the largest unstable modes of fragmentation within the filament.}
\label{fig_cases}
\end{figure*}

In the upper left panel of Figure \ref{fig_cases}, we show the obtained CMF within non-magnetized filaments of MpL 40, 160, 640 ~$\Mspc$ with solid curves. 
The filament properties are derived from equations (\ref{eq_vir_cyn}) and (\ref{eq_Ma}), 
and the mass functions are calculated with equations (\ref{eq_MF})-(\ref{eq_dddl}).
The number of fragments increases with increasing density inside the filament, 
and the characteristic mass decreases accordingly. 
Two levels of magnetization are shown for the 160 $\Mspc$ filament, namely $\mathcal{M}_\alfv = 2$ (light green dashed) and $1$ (dot-dashed), which
correspond to $E_\mathrm{mag} / E_\mathrm{nt} = 0.1$ and $0.5$, respectively. 
The presence of magnetic field not only provides support against gravity, but also narrows the spectrum of density fluctuations, as equation (\ref{eq_sig0}) now becomes, for $\gamma_\magb = 0.5$ \citep{Molina12},
\begin{equation}
\sigma_0^2 = \log(1+ 3b^2\mathcal{M}_\mathrm{1D}^2{\beta_0 \over \beta_0 +1}),  
\label{eq_sig_mag}
\end{equation}
where $\beta_0 = 2c_\sound^2/{v_\alfv^\ast}^2$ is the ratio between thermal and magnetic pressure.
We note that with increasing magnetic field intensity
the fragmentation is significantly suppressed and small cores are prevented from collapsing. 
The 640 $\Mspc$ filament with $\mathcal{M}_\alfv = 1$ (dark green dot-dashed curve) shows that, at given $\mathcal{M}_\alfv$, the characteristic mass for fragmentation {\it increases} with MpL, in contrast to the hydro case. 
We also note that, in both cases, massive cores form dominantly inside filaments with the highest MpL. 
This is in line with the finding by \citet{Li16} from the ATLASGAL survey. 
Note that our model works well only with filaments that are sufficiently supercritical/turbulent ($40~\Mspc$ appears to be largely sufficient with respect to critical value of $30$) since it is based upon the statistical properties of the turbulent medium. Filaments close to criticality have less fluctuations that induce local collapse, and a direct application of the present model produces fragments that are too large and the results quickly diverges when thermal support is dominant. This is easily shown with simple virial calculation of a thermally supported ellipsoid, of which the size quickly increases with decreasing MpL ($\sim 1~\pc$ fragments within a filament of $30~\Mspc$), as will be shown in Section \ref{sec_high_mass_end}. The studies of these filaments require detailed stability analysis \citep[e.g.][]{Stodolkiewicz63,Inutsuka92}.

%\begin{figure}[]
%\centering
%\includegraphics[trim=0 10 0 0,clip,width=0.5\textwidth]{CMF_cases_l.pdf}
%\caption{Number distribution as function of the clump size in different filaments. Same cases are plotted as those in Figure \ref{fig_cases}. All filaments have 0.1 pc diameter and 10 pc length The quality $dN/d\log{l}$ is plotted against the major axis $2l$ of the self-gravitating clumps that are destined to collapse. The sizes before collapse could be regarded as the modes of fragmentation.}
%\label{fig_cases_l}
%\end{figure}

In the upper right panel of Figure \ref{fig_cases}, we display the number of individual cores as function of their size just at the edge of collapse. 
This is the typical size at which the virial equilibrium condition is obtained, with 
unstable clumps becoming ultimately a core or a group of cores, depending on their size, as will be discussed in Section \ref{st_Hierarchical}. 
We stress that this reflects the size of the unstable density fluctuations \emph{before} collapse takes place. It thus should be interpreted as a local fragmentation scale rather than the final size of the cores at the end of the collapse. 
In the absence of magnetic field, the typical core fragmentation length scale decreases with increasing MpL, because of both  the decreasing Jeans mass and the increasing level of turbulence. 
In contrast, in the magnetized case, under energy equilibrium condition,
the fragmentation scale increases at fixed MpL with increasing magnetic support, 
illustrating the fact that magnetic field hampers fragmentation.

The following calculations illustrate the behavior of the CMF under specific filament conditions and scaling laws for density dispersion, turbulence, and magnetic field. They give physically comprehensive results to shed light on the impact of filamentary geometry on core formation/distribution. More realistic calculations could be done by a direct numerical application of the scaling relations, if available from observations or theories. This model provides a framework for fragmentation of filamentary structures. While we consider only star-forming filaments of sub-parsec width, MpL  $> 16~\Ms$, and visual extinction $A_\mathrm{v} \gtrsim 7$ or column density $>7\times 10^{21}~\cc$ \citep{Heiderman10, Lada10, Andre10, Andre14, Konyves15}, this model can be readily applied to fragmentation of filaments at all scales provided the corresponding conditions and scaling laws are used. 
%In Section \ref{sec_v}, we show the output of the simulation of a filament fragmentation to illustrate the process.

\subsection{Low-mass end and peak position}
\label{st_M_peak}
The peak of the CMF occurs at $l < R$, 
where the turbulent rms velocity is comparable to the thermal sound speed. 
For sake of simplicity, we thus neglect the turbulence and the second term inside the parenthesis of equation (\ref{eq_MF}), since it does not contribute much at small scales, to derive analytically the peak position of the CMF. 
With purely thermal support and spherical geometry ($r\sim l$), we obtain from the time-dependent CMF
\begin{eqnarray}
\widetilde{M} &=& \Lambda^{-1}\widetilde{l}, ~\text{and}\\
\mathcal{N} &=& {m_\lambda \over M_0^2}{2 \Lambda^3 \over \sqrt{2\pi \sigma^2}} \exp{\left(-{\sigma^2\over8}\right)} \left(\widetilde{M}\Lambda^{3\over2}\right)^{-4-{2\log{\left(\widetilde{M}\Lambda^{3\over2}\right)}\over\sigma^2}} . 
\end{eqnarray}
This simply reproduces the result of \citet{HC08, HC09, HC13}.
The low-mass end exhibits an exponential cut-off, as found in other studies \citep{Padoan97}. 
The peak mass corresponds to 
\begin{equation}
\widetilde{M}_\mathrm{peak} = \Lambda^{-{3\over2}} \exp{(-\sigma^2)}, 
\end{equation}
where $d\mathcal{N}/dM = 0$. 
Since $\Lambda \propto m_\lambda$ and $\exp{(\sigma^2)} = 1+3b^2\mathcal{M}_\mathrm{1D}^2 \simeq 1+b^2(m_\lambda -m_\crit)/ m_\crit$, 
the peak mass scales as 
\begin{equation}
M_\mathrm{peak} = M_0~ \widetilde{M}_\mathrm{peak} \propto m_\lambda^{1-{3\over 2} -1} \propto m_\lambda^{-{3\over2}}.
\end{equation} 
This scaling is only valid for supercritical filaments, i.e. under the condition that the filaments MpL is sufficiently larger than the critical value. 
%This is confirmed by Figure \ref{fig_cases}, 
%where the 30 $\Mspc$ case slightly deviates from this relation since it is only marginally supercritical. 

As discussed above, magnetic field seems to play a major role in filament fragmentation, 
as indeed suggested by numerous observational results \citep[e.g.][]{Li15, Zhang15, Pillai15}. 
Moreover, filaments by themselves are very often the product of gravo-turbulent collapse guided by magnetic fields. 
Especially at small scales and high densities, magnetic field tends to be very efficient in preventing fragmentation. 
When $\gamma_\magb = 0.5$, magnetic support behaves like an isothermal gas. 
This allows us to refine the above discussion by simply adding the magnetic term, such that
\begin{equation}
\widetilde{M} = \Lambda^{-1}\widetilde{l} \left( 1 +{V_\alfv^2\over 6}  \right).
\end{equation}
The peak mass becomes
\begin{equation}
\widetilde{M}_\mathrm{peak} = \Lambda^{-{3\over2}} \left( 1 +{V_\alfv^2\over 6}  \right)^{{3\over2}} \exp{(-\sigma^2)}.
\end{equation}
If we assume some energy equipartition such that the magnetic energy is proportional to the turbulent energy, 
$\mathcal{M}_\alfv$ is a constant and $V_\alfv^2 \propto m_\lambda/m_\crit$.
At the same time, the density PDF width is modified in the presence of magnetic field. 
Following \citet{Molina12}, we take
\begin{eqnarray}
\sigma^2 &=& \log(1+ 3b^2\mathcal{M}_\mathrm{1D}^2{\beta_0 \over \beta_0 +1})  \\
&\approx&~ \log(6b^2\mathcal{M}_\mathrm{1D}^2/ V_\alfv^2) ~=~ \log(2b^2\mathcal{M}_\alfv^2), \nonumber
\end{eqnarray}
where $\beta_0 = 2c_\sound^2 / {v^\ast_\alfv}^2 = 2/V_\alfv^2$. 
The narrowing of the PDF in the magnetized cases depends on the ratio between thermal and magnetic pressure $\beta_0$ .
The approximation is done under the condition that both turbulent and magnetic pressures are much larger than the thermal pressure, i.e., $\mathcal{M}_\mathrm{1D} \gg 1$ and $\beta_0 \ll 1$. 
 In the presence of magnetic field with constant $\mathcal{M}_\alfv$, the peak mass then now depends on the filament MpL as
\begin{equation}
\label{eq_ml_1}
M_\mathrm{peak} = M_0 ~\widetilde{M}_\mathrm{peak} \propto m_\lambda^{1-{3\over 2}+{3\over 2}-0} \propto m_\lambda
\end{equation} 
This relation clearly illustrates how magnetic field increases the typical size of fragmentation in a filament. 
If the magnetic field obeys more complicated relations, with $\gamma_\magb \ne 0.5$ or a fraction of energy equipartition which varies with MpL, 
this above derived peak position dependence will be affected accordingly, but the general behavior should remain unchanged. 
For a value $\gamma_\magb > 0.5 $, the magnetic field prevents the fragmentation of high-density gas and limits the low mass part of the CMF without affecting substantially the peak position. 
On the contrary, magnetic fields characterized by a value $\gamma_\magb < 0.5$ narrow less significantly the width of the density PDF and should result in a CMF between the hydro and the $\gamma_\magb = 0.5$ cases \citep{Molina12}. 
 Note that in our model the magnetic field provides mechanical support as a pressure term; similar pressure support could also be due to other mechanisms such as cosmic rays or stellar radiative feedback.

\subsection{High-mass end}
\label{sec_high_mass_end}
In the filamentary topological configuration, 
the high-mass end of the CMF enters the one-dimension domain for density fluctuations, 
with fundamental differences compared with the 3D spherical case. 
As shown in \citet{HC08}, the turbulent support is dominant at the high-mass end of the CMF\footnote{As described in \citet{CH11}, we stress that the turbulent support should not be understood as a static, pressure-like support, but rather as a statistical, dynamical process picking up regions dense enough to collapse in a {\it turbulent} medium}.
This allows us to discuss the 3D to 1D transition with simplifying assumptions, 
by dropping the thermal and magnetic terms in equation (\ref{eq_M}), 
so that the critical collapsing density reads
$\widetilde{\rho} \propto \mathcal{M}_\mathrm{1D}^2 \widetilde{l}^{2\eta_\turb+1} / (\widetilde{r}^2\widetilde{l}w_g(\eta))$. 
This quantity decreases with increasing scale in the spherical regime, where $\widetilde{r} \sim \widetilde{l}$, 
while it increases with scale when the overdense fluctuations becomes significantly prolate, 
i.e., $\widetilde{l} \gtrsim1$ and $\widetilde{r} \sim 1$. 
The result is a sign change of the first term in brackets in equation (\ref{eq_MF}) from positive to negative, 
that imposes an {\it upper mass limit of fragments inside the filament}.  
Assuming isothermality and ignoring magnetic field, 
we evaluate this cutoff by calculating the size $\widetilde{l}_\mathrm{M}$ at which the density derivative is zero: 
\begin{eqnarray}
{d\delta \over d \widetilde{l}} &=& {1\over \widetilde{\rho}}{d\widetilde{\rho} \over d \widetilde{l}} =0 ~\text{ in equation (\ref{eq_MF}),} \\
{d\widetilde{\rho} \over d \widetilde{l}} &=& {\Lambda^{-1} \over w_\grav \widetilde{r}^2}
 \left[  2\eta_\turb\mathcal{M}_\mathrm{1D}^2\widetilde{l}^{2\eta_\turb-1} - \right. \\
&&\left. {1\over\widetilde{r}}\left({w_\grav(\eta)^\prime \over w_\grav(\eta)}\widetilde{r}^n+2\eta^{-(1+n)}\right)\left(1+\mathcal{M}_\mathrm{1D}^2\widetilde{l}^{2\eta_\turb}\right)\right].\nonumber
\label{eq_dd0}
\end{eqnarray} 
The asymptotic value of $\widetilde{l}_\mathrm{M}$ at large Mach is 2.5 for $n=1$ and 1.75 for $n=2$, 
and it decreases with $n$ and approaches 1 when $n \rightarrow \infty$. 
The smoothing of the geometrical transition allows the collapse of fragments at size slightly larger than the filament width. 
The corresponding cutoff mass is calculated using equation (\ref{eq_M}) 
\begin{figure}[]
\centering
\includegraphics[trim=0 0 0 0,clip,width=0.5\textwidth]{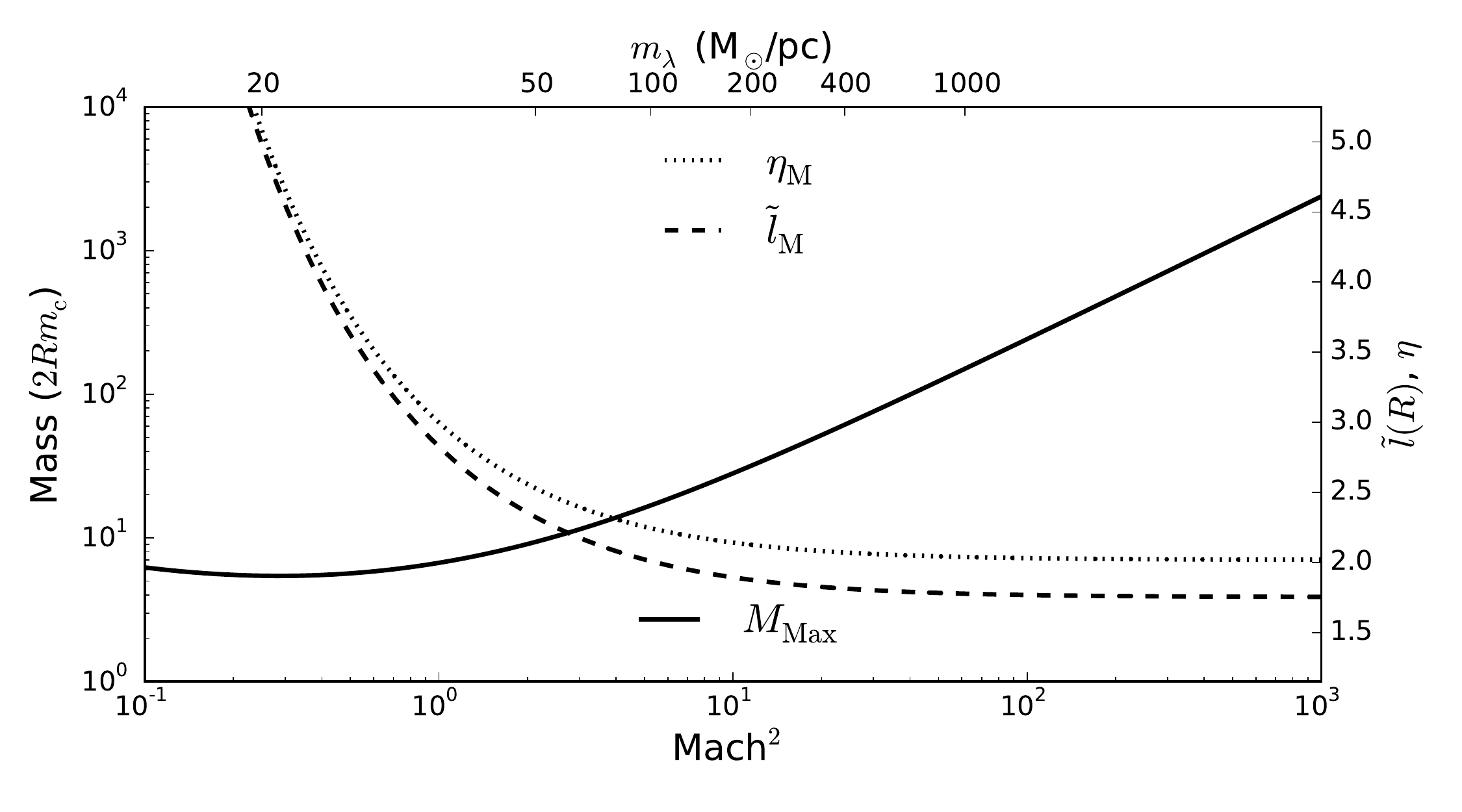}
\caption{Mass of the largest fragment (solid curve,  in units of  the characteristic mass   of the critical filament  $2R m_\crit$, MpL multiplied by the diameter) as function of the Mach number square, under isothermal (10 K) and non-magnetized assumptions. Also plotted are the corresponding size (dashed curve, normalized to the filament width) and aspect ratio (dotted curve). The smoothing parameter $n=2$ is used. The corresponding MpL in virial equilibrium calculated from equation (\ref{eq_vir_cyn}) is also shown (top axis).}
\label{fig_Mmax}
\end{figure}
\begin{eqnarray}
M_\mathrm{Max} &=& {4\over3}m_\lambda R\widetilde{M}(\widetilde{l}_\mathrm{M})  \\
& =& {5 \over 2}m_\crit R {\widetilde{l}_\mathrm{M}\over w_\grav(\eta_\mathrm{M})} 
\left(1+\mathcal{M}_\mathrm{1D}^2\widetilde{l}_\mathrm{M}^{2\eta_\turb}\right) \nonumber\\
&\approx& 2.4 ~(2m_\crit R) ~\mathcal{M}_\mathrm{1D}^2 \nonumber\\
&\approx& 3.8~\Ms \mathcal{M}_\mathrm{1D}^2 \left({T \over 10~\mathrm{K}}\right)\left({R \over 0.1~\pc}\right). \nonumber
\end{eqnarray} 
This approximation is only valid when the Mach number is sufficiently large. 
This relation, with masses normalized to $2Rm_\crit$,  the characteristic mass of the critical filament , is plotted as function of $\mathcal{M}_\mathrm{1D}^2$ in Figure \ref{fig_Mmax}. 
The normalization value is about $1.6 ~\Ms$ for an isothermal critical filament at 10 K of $0.1~\pc$ diameter. 
We caution that the above derived value is an upper limit since the second term in equation (\ref{eq_MF}) is negative, 
and will slightly decrease the value. 
This result imposes an upper limit to the most massive star that can form inside filaments. 
Since the square of the Mach number is roughly proportional to the MpL, 
we need a filament 10 times more massive than the thermally critical value ($16 ~\Mspc$ at 10 K) to form a $30 ~\Ms$ star. 
The transonic and subsonic filaments are predominantly thermally supported and we can see from Figure \ref{fig_Mmax} that the geometrical upper cutoff goes to infinity for small Mach numbers. 
The lack of low and intermediate mass fragments inside these filaments thus stems from the lack of large-scale turbulent induced fluctuations \citep{HC08, CH11}, yielding a very steep high-mass slope, rather than from the presently discussed geometrical limitation in massive (supersonic) filaments. 
These low MpL filaments tend to fragment mostly as in the standard Jeans gravitational scenario \citep{Hacar13,Kainulainen16a}.

In magnetized cases with $\gamma_\magb = 0.5$, 
one can simply replace the sonic Mach $\mathcal{M}_\mathrm{1D}^2$ number by the Alfv\'en-sonic Mach number $\mathcal{M}_\mathrm{1D}^2\beta_0 / (1+\beta_0)$ in equation (\ref{eq_dd0}) and multiply the resulting mass by $(1+\beta_0)/\beta_0$. 
We note from Figure \ref{fig_Mmax} that the maximal size of fragment $\widetilde{l}_\mathrm{M}$ is always between 2 and 3 for filaments with $\mathcal{M}>1$, while it increases to infinity for smaller $\mathcal{M}$.
At this scale, the turbulent and magnetic field fluctuations with respect to the mean fields are small enough so that the maximal fragment mass almost solely depends on the MpL of the filament. 
This is confirmed in Figure \ref{fig_cases}, which shows that the mass cutoff for the same MpL and different magnetic energy levels varies by less than a factor 2. 
%In the low $\mathcal{M}$ regime, characteristic of filaments close to pure thermal support, the observational lack of high mass cores is due to the lack of large-scale density fluctuations \citep{HC08, CH11} and is thus of statistical rather than geometrical origin, and their behavior is not well predicted with our model.

\subsection{Hierarchical fragmentation: The mass function of groups of cores.}
\label{st_Hierarchical}
Observations show that massive filaments fragment into clumps at typical separations a few times the filament width, i.e. the filament typical Jeans length. 
These clumps further break up into small groups of cores. 
Such a hierarchical fragmentation has been reported in the Taurus filament L1495/B213 \citep{Hacar13}, IRDC G11 \citep{Kainulainen13}, 
the Orion complex \citep{Takahashi13}, 
and in the Spokes cluster \citep{Pineda13}. 
\citet{Teixeira16} found from observations of the Orion Molecular Cloud 1 northern filament that there are two typical separations between paired dense cores at 0.06 and 0.01 pc (projected on the plane of sky), respectively,
suggesting that the filament has undergone fragmentation into small groups. 
\citet{Kainulainen17} also found groups of fragments separated by typically 0.3 pc in the Orion A integral shaped filament. 

Such a 2-level fragmentation process naturally derives from our formalism. 
Indeed, returning to Section \ref{st_HC}, 
we see that, 
considering the aforementioned group of cores as the {\it largest} self-gravitating fluctuations in a filament, 
the total mass in these structures is given by
\begin{equation}
M_\mathrm{tot}(l) =  L_\mathrm{tot} \int\limits_{M_l^\crit}^\infty M^\prime  \mathcal{N}_\crit^{m_\mathrm{\lambda}}(M^\prime) P(M_l^\crit | M^\prime) dM^\prime,
\label{eq_int_mass_2}
\end{equation}
where the integral now must be taken from $M \ge M_l^\crit$ to infinity and now $P(M^\prime |M_l^\crit)=1/2$. 
This would correspond to a {\it first}-crossing condition in the excursion set formalism \citep{Hopkins12a, Hopkins12b, Hopkins13a}. 
The mass function of these group of cores is then calculated exactly as described in Section \ref{st_HC}  for the CMF. 
When considering the cloud-in-cloud problem, 
the probability that a mass at scale $l$ exceeds the critical density inside a larger scale $l^\prime$ corresponding to $M^\prime$ is $1/2$ if the density threshold is constant \citep{Yano96, HC08}. 
In our formalism, the density threshold for collapse is given by
the virial condition and thus this probability is less clear. 
However, it should be roughly constant and thus should not affect the shape of the resulting mass function. 
We thus keep the value $1/2$.  
The above considerations thus simply result in a change of sign and an extra factor 2 in equation (\ref{eq_MF}). 

The group mass function for different filaments is shown in the left lower panel of Figure \ref{fig_cases}. 
In this regime, the second term becomes less negligible compared to the first term in equation (\ref{eq_MF}); this is discussed in Appendix \ref{appen_g}. 
As shown in Figure \ref{fig_groups}, the mass function in the (large-scale) 1D domain is a strictly decreasing function of $M$ and $l$. 
The very steep increase at the low-mass end arises from the geometrical smoothing. 
The mass function is monotonically decreasing with $l$ for $n \rightarrow \infty$,  if a cylinder with no envelope and no geometrical smoothing is considered .
This means that the dominant mode of fragmentation occurs at intervals at least $2\widetilde{l}_\mathrm{M}$,  the major-axis of the smallest unstable first-crossing ellipsoid . Recall that $\widetilde{l}_\mathrm{M} = 1.75$ for $n=2$ (Section \ref{sec_high_mass_end});  both $\eta_\mathrm{M}$ and $\widetilde{l}_\mathrm{M}$ tend to 2 if no geometrical smoothing is considered, and the fragmentation is purely 1D, i.e. $n \rightarrow \infty$ . 

It is obvious that groups of cores cannot have sizes initially smaller than the filament width, 
and the most prominent scale is a few times the filament width, as discussed in Section \ref{sec_high_mass_end}. 
Groups of cores are products of filament fragmentation along the longitudinal direction, 
as found in 1D perturbation analysis of filaments. 
When turbulence is large, i.e., for high MpL and high $\mathcal{M}_\alfv$, the distribution becomes wider, as anticipated from equations (\ref{eq_sig}) and (\ref{eq_sig0}).

The black line in Figure \ref{fig_cases} corresponds to the Salpeter IMF, $M^{-1.33}$. 
We see that in the most massive filaments the shape of the group mass function approaches this limit, 
while it becomes steeper with decreasing MpL or increasing magnetization. 
This means that in filaments with lower MpL or larger magnetic field, 
the separation between groups of cores is more regular, 
and large variations are only expected in filaments with high MpL and low magnetic field, i.e., high levels of turbulence. 
We note that the second term in the group mass function (equation (\ref{eq_MF}) modified with equation (\ref{eq_int_mass_2})) corresponds to the distribution derived by  \citet{Inutsuka01}. 
Our results, however, differ from this approach by the fact that we use the virial equilibrium to define the collapse condition, while a constant density contrast threshold is assumed in \citet{Inutsuka01}. 
The model presented in this paper thus shows that 1D perturbative studies of filaments can only yield fragmentation into {\it groups} of cores, and do not describe the sub-fragmentation process leading to the core mass function itself.

\subsection{Visualization of self-gravitating objects in simulated filaments}
\label{sec_v}
To yield a more comprehensive view of our model and illustrate the fragmentation process, we simulate filaments harboring density fluctuations and identify the self-gravitating objects within these filaments. The simulation is performed as follows: first, we produce a random gaussian field of log-density with a (3D) power spectrum index $-11/3$ on a $1024^3$ grid. Second, a filamentary region is arbitrarily extracted (we chose 20 grid points for the radius and 800 points in length) and the fluctuations are renormalized such that the dispersion corresponds to equation (\ref{eq_sig_mag}) for a virialized filament, for given $m_\lambda$, $\mathcal{M}_\alfv$, and $\mathcal{M}$ deduced from equation (\ref{eq_vir_cyn}). With turbulent and Afv\'en velocities characterized respectively by equations (\ref{eq_vturb}) and (\ref{eq_vmag}), we identify self-gravitating objects that satisfy the virial condition (\ref{eq_vir}). The first-crossing (groups) or last-crossing (cores) unstable structures are identified by progressively decreasing or increasing the scale $l$ when looking at all eligible ellipsoidal objects defined by equation (\ref{eq_r_of_l}). Everytime a self-gravitating object is detected, it is removed before proceeding further with the next $l$ value. We show a few examples in Figure \ref{fig_visu_fila} for $m_\lambda = 400 ~\Mspc$ and $\mathcal{M}_\alfv =20$ (cases 1, 2, 3, and 5) or $5$ (cases 4 and 6). The last two pairs of results are generated from the same field of density fluctuations, and only differ in $\mathcal{M}_\alfv$. It is clear that the magnetic field hampers the fragmentation. Note that it is computively expensive to reach large statistics with such simulations since the dense regions make up only a small volume fraction of the entire filament. 
\begin{figure*}[]
\centering
\includegraphics[trim=50 0 50 0,clip,width=\textwidth]{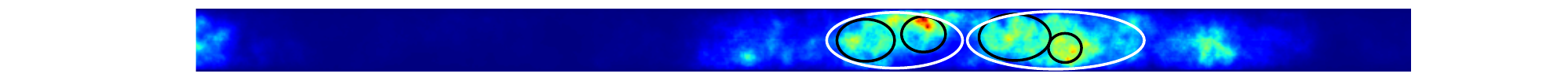}
\includegraphics[trim=50 0 50 0,clip,width=\textwidth]{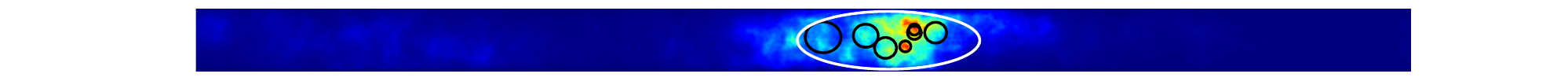}
\includegraphics[trim=50 0 50 0,clip,width=\textwidth]{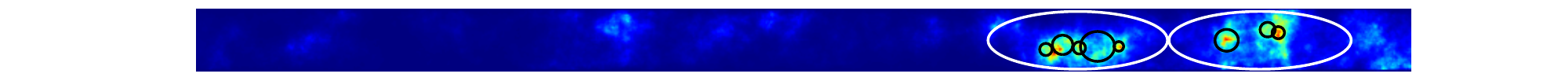}
\includegraphics[trim=50 0 50 0,clip,width=\textwidth]{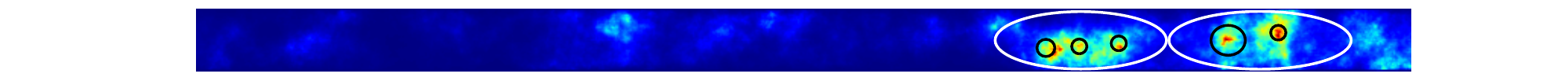}
\includegraphics[trim=50 0 50 0,clip,width=\textwidth]{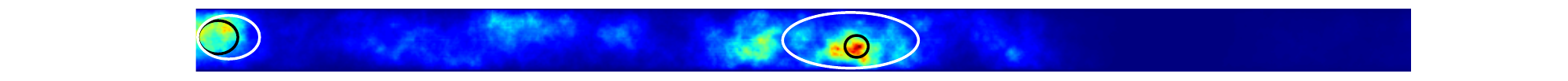}
\includegraphics[trim=50 0 50 0,clip,width=\textwidth]{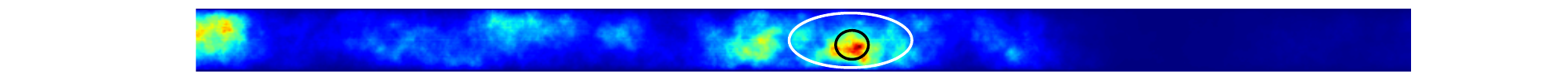}
\caption{Visualization of a filament fragmentation for $m_\lambda = 400 ~\Mspc$ and $\mathcal{M}_\alfv =20$ (cases 1, 2, 3, 5) or $5$ (cases 4, 6) with 4 different randomly generated lognormal density fields, radius resolved by 20 points. Column density is shown with color scale. The first-crossing (groups) is marked with white contours, and last-crossing (cores) in black. Cases 3 and 4 have the same density field with different $\mathcal{M}_\alfv$ to illustrate the effect of magnetization; same cases 5 and 6.}
\label{fig_visu_fila}
\end{figure*}

\subsection{Caveats for filament analyses}
We have treated the core formation inside a filament in global equilibrium. The formation of filament and cores, however, is not a static process. The time dependence introduced by equation (\ref{eq_time_dep}) physically reflects the fact that accretion onto the filament constantly replenishes it with material and density fluctuations. This corresponds to a quasi-stationary perspective of the filament, such that its evolution timescale is longer than that of core formation, and multiple generations of cores can form inside a filament. Our model provides an instantaneous view of filament fragmentation given some characteristic conditions, which could be slowly-varying. This corresponds to a snapshot of a filament population, as seen in observations. 
Note also that we only studied the last level of structure formation, namely the fragmentation of filaments into prestellar cores. The formation and the properties of filaments themselves remains an issue to be better understood. Nevertheless, even though filaments in this study are admittedly idealized, this study brings relatively robust results. First, the geometry imposes an upper limit for the core mass which is about a few times the filament mass per unit length times the width. Second, the number of filaments with high MpL is small, which implies a small number of massive stars in isolated filaments. This in turn suggests that massive stars should form preferentially in specific environments such as filament nests or dense clusters. Our model does not address so far this issue. 

 As mentioned earlier, it should also be bore in mind that, for sake of simplicity, we do not consider any MHD dynamics but treat the magnetic field as an isothermal, isotropic pressure term (ignoring turbulent fluctuations) in both the criterion for gravitational collapse and the expression for the magnitude of density fluctuations. Any other source of pressure (e.g. higher gas thermal pressure or additional pressure from radiation or cosmic rays) will behave similarly. Further studies are required to better understand the dynamical effects (field lines guiding the flow of the gas) and the turbulent fluctuations of the magnetic field.  
 
%________________________________________________________________________________
\section{CMF from a distribution of filaments}
\label{st_fila_pop}
\subsection{The convolution: A two-mode and two-step process}\label{st_conv_mod}
Observations have revealed that there is a distribution of filaments with different column densities, while their width varies very weakly \citep{Arzoumanian11}. The question we want to address now is the consequence of this filament distribution on the final CMF.
Indeed, in the presence of density anisotropies, which is usually the case, 
a gravitationally unstable massive clump tends to collapse along its short axis. 
Small gravitationally unstable clumps should thus collapse directly into prestellar cores, 
whereas large ones will collapse into filaments, 
that will further fragment into cores. 
We thus suggest such a 2-mode process for prestellar core (and thus star) formation. 
The filamentary mode of core formation comprises two steps: filament formation (see Appendix \ref{appen_y} for a simple model that suggests a formation mechanism for the filament distribution) and then core formation.

Remembering that $\mathcal{N}_\mathrm{c}^{m_\mathrm{\lambda}}(M)$ denotes the CMF inside a filament of MpL $m_\mathrm{\lambda}$, 
the total mass in collapsing cores inside the whole system composed of various filaments is given by
\begin{eqnarray}
\label{eq_coll}
M_\mathrm{coll}& =& \int\limits dM_\mathrm{sys}  \\
&=& \int\limits M \mathcal{N}_\mathrm{sys}(M) ~dM \nonumber\\
&=& \int\limits L^{m_\mathrm{\lambda}} \int\limits M \mathcal{N}_\mathrm{c}^{m_\mathrm{\lambda}}(M) dM ~~ \mathcal{N}_\mathrm{f}(m_\mathrm{\lambda}) ~dm_\mathrm{\lambda} \nonumber\\
&=& \int\limits M \left( \int\limits L^{m_\mathrm{\lambda}} \mathcal{N}_\mathrm{c}^{m_\mathrm{\lambda}}(M) \mathcal{N}_\mathrm{f}(m_\mathrm{\lambda}) dm_\mathrm{\lambda} \right) dM, \nonumber
\end{eqnarray}
where $\mathcal{N}_\mathrm{sys}$ is the convolved CMF of the entire system, 
$L^{m_\mathrm{\lambda}}$ is the filament length (assumed to be a function of the  MpL), 
and $\mathcal{N}_\mathrm{f}$ is the filament MpL function (FMLF), in number per MpL. 
The resulting (two-step) CMF for large unstable clumps thus reads
\begin{equation}
\label{eq_conv}
\mathcal{N}_\mathrm{sys}(M) = \int\limits L^{m_\lambda}  \mathcal{N}_\mathrm{c}^{m_\lambda}(M) \mathcal{N}_\mathrm{f}(m_\lambda) dm_\lambda, 
\end{equation}
which is a convolution of the FMLF and the CMF inside individual filaments. 
By integrating over the total filament population, characterized by the MpL, 
we obtain the CMF of the whole system, 
which is arbitrarily normalized since we do not know the filament number density. 
The above equation requires a knowledge of the filament population, 
i.e., the distribution as function of MpL and the corresponding length,
as well as how the CMF inside a filament depends on the filament properties. 

In order to handle this task and to examine how the system CMF depends on these filament parameters,
we perform a simple parametric analysis. 
We first introduce the dependence of the filament properties upon MpL such that
\begin{eqnarray}
L^{m_\mathrm{\lambda}}  &\propto& m_\mathrm{\lambda}^\Gamma \label{eq_L}\\
\mathcal{N}_\mathrm{c}^{m_\mathrm{\lambda}}(M) &\propto&  m_\lambda^{1-\beta} ~\delta{(M-m_\mathrm{\lambda}^\beta)} \label{eq_delta}\\
\mathcal{N}_\mathrm{f}(m_\mathrm{\lambda}) &\propto& m_\mathrm{\lambda}^{-\alpha}. \label{eq_mf}
\end{eqnarray}
Equation (\ref{eq_L}) means that the length of the filament scales with MpL, 
with $\Gamma=0$ if all filaments have the same $L^{m_\mathrm{\lambda}}=$ constant. 
If a clump follows the Larson's relation $M \propto R^2$ and forms a filament without contracting in the longitudinal direction, then $m_\lambda \propto M/R \propto R$ and $\Gamma=1$. 
Equation (\ref{eq_delta}) is a simplifying assumption that the filament regularly fragments into pieces of the same mass, that is governed by the MpL. 
If all filaments have the same width and fragment into roughly spherical pieces, $\beta \approx 1$. 
If denser filaments fragment into smaller pieces, then $\beta < 1$, and vice versa. 
The number of fragments is proportional to $m_\lambda^{1-\beta}$ if all the filament mass goes into fragments. 
In reality, the CMF inside filament should be a broad function peaked around this value, as discussed in the previous section. 
Observations \citep[][Arzoumanian et al. in prep]{Andre14} suggest a value $\alpha \simeq 2.2$ for the MpL function (\ref{eq_mf}). 
With the scaling relations in equations (\ref{eq_L})-(\ref{eq_mf}), integration of equation (\ref{eq_conv}) yields for the convolved CMF
\begin{equation}
\mathcal{N}_\mathrm{sys}(M) \propto M^{{\Gamma-\alpha+1\over \beta}-1}.
\label{eq_N_mod}
\end{equation}
Assuming $\beta=1$ (see equation (\ref{eq_ml_1})), we obtain a powerlaw exponent -2.2 for $\Gamma=0$ and -1.2 for $\Gamma=1$. This illustrates the dependence of the CMF upon the filament MpL. 
Note that $\Gamma$ and $\alpha$ are somehow degenerate parameters.
The physical interpretation is simply that one long filament and two shorter filaments with the same total length contribute equally to the CMF since, according to our model, the filament length does not play an important role in the fragmentation process.
Variations of $\beta$ around unity slightly changes the slope. 
Smaller $\beta$ results in steeper slopes, and vice versa. 
From Section \ref{st_M_peak}, we see that $\beta$ is probably close to 1 if the filaments are magnetized. 
On the other hand, if the CMF inside individual filaments has itself a powerlaw tail, 
the convolved slope of the integral CMF would be the shallowest one between this slope and the afore-derived one.

\subsection{The filament population}\label{st_filpop}
As mentioned earlier, this paper aims at studying fragmentation at the filament scale. The formation of a filament population from hierarchical condensation of large scale structures is beyond the scope of this study.
We thus take the filament distribution inferred from observations to characterize the behavior of the final CMF.
Some statistical filament properties are not readily available from observations, 
such as the length, the MpL distribution, and the strength of the magnetic field, and are thus subject to large uncertainties. 
In contrast, the filament width seems to always be close to 0.1 pc, regardless of the MpL \citep{Arzoumanian11, Arzoumanian13,Koch15}. 
Simulations \citep{Federrath16} and analytical models \citep{HA13, Auddy16} have been proposed to explain the universality of the observed width. 
Some observations suggest that the filament width may increase with MpL \citep{Hill11}. 
Increasing the filament width introduces roughly a linear scaling to the resulting mass of fragments. 
This is a small effect, and discussed in Appendix \ref{appen_R}. 
We will use the 0.1 pc value as the fiducial filament diameter in our model,  but this model can be readily applied to filaments of different widths.  

The determination of the length of filaments largely depends on the criteria and algorithms used in the observational analysis. 
One long filament with large fluctuations could be regarded as several shorter filaments if there is some small density break along the filament. 
Since no strong correlation has been observed between the MpL and the length of filaments \citep{Konyves15}\footnote{The column density PDF from the filamentary region has almost the same slope of the FMLF distribution, reflecting that the total length of filaments at one given MpL is always the same.}, and since, as just shown, the filament length does not have a strong impact on the CMF, we simply assume uniform filament lengths, independent of MpL, keeping in mind that in reality the length probably has some dispersion around a mean value. 
For our fiducial FMLF, we take $dN/dm_\lambda \propto m_\lambda^{-2.2}$ \citep{Andre14}. 
We choose these canonical values to illustrate the bound core mass distribution produced by fragmentation of a filament population. 
However, if either the filament length depends on MpL or if the FMLF has a different slope, 
the behavior of the convolved CMF for the whole filament population can be derived from equations (\ref{eq_coll}) and (\ref{eq_conv}) with the appropriate relations.
Note that although the density PDF inside individual filaments is of lognormal form, 
the integration over a powerlaw FMLF results in a powerlaw PDF for the whole filamentary region. 

The role of magnetic field is even less well understood. 
While most filaments are observed to be threaded by magnetic fields, 
the strength of the field is highly debated. 
We thus explore different levels of magnetization.

%________________________________________________________________________________

\subsection{The resulting CMF and its dependence upon the filament various properties}
\label{st_conv}
In this section, we apply our calculations to different filament MpL distributions and magnetic field strengths. 
Note that we do not add any time weighting between filaments of various MpL, 
as that has been done for self-gravitating cores. 
This simplifying assumption is used here as the filaments do not seem to be radially collapsing in one free-fall time, 
while we stress that the formation and evolution of filaments remain to be clarified.

\subsubsection{Non-magnetized case}
\label{st_conv_hydro}
\begin{figure}[]
\centering
\includegraphics[trim=0 0 0 0,clip,width=0.5\textwidth]{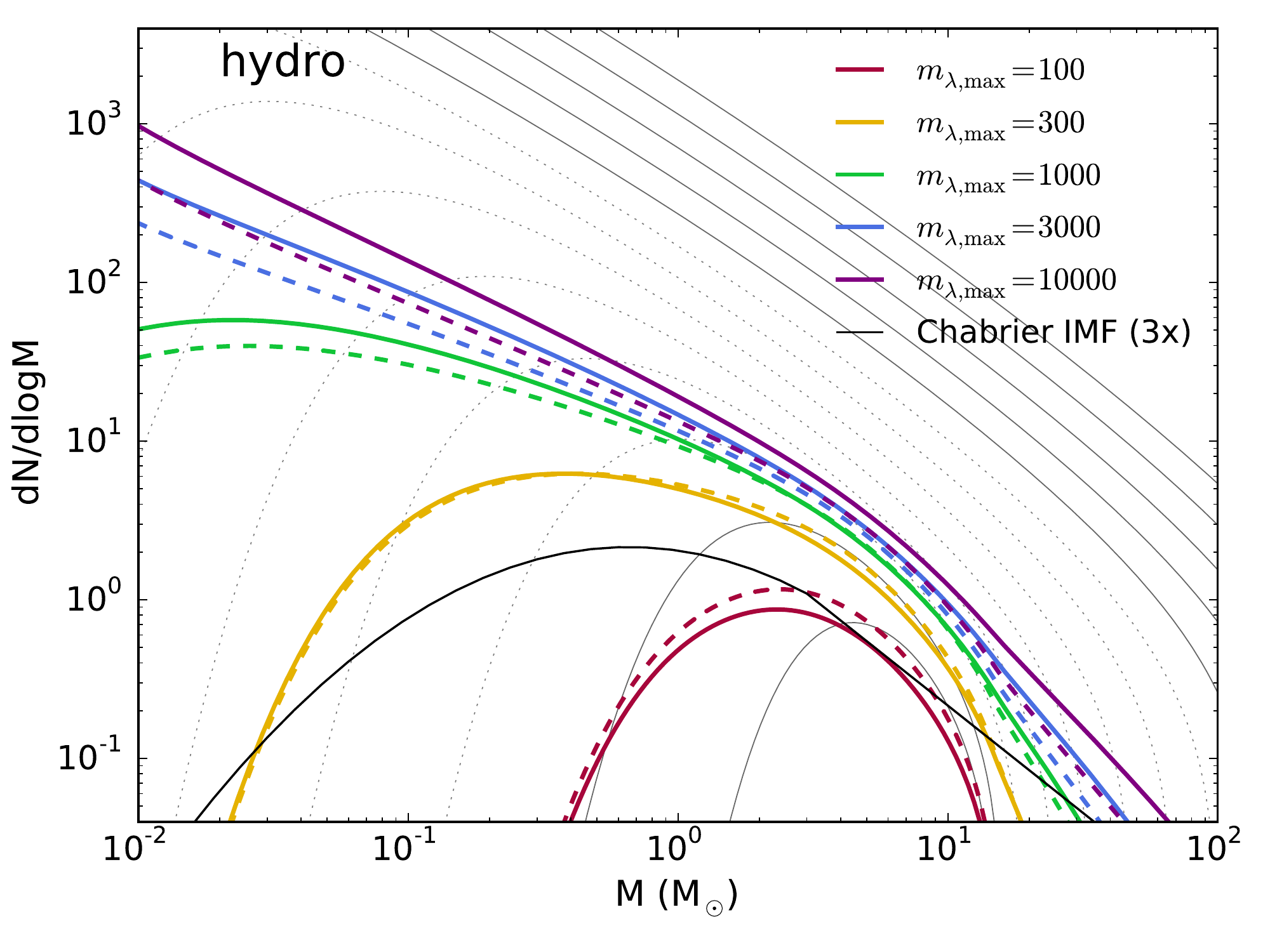}
\caption{Convolved system CMF for non-magnetized filaments (as obtained from equation (\ref{eq_conv})). CMF per unit length of individual filaments are plotted in solid dark grey curves ($<100~\Mspc$ and $>1000~\Mspc$) and dotted light grey curves ($100~\Mspc <$ MpL $<1000~\Mspc$), with MpL increasing from bottom to top. Convolutions of different populations are shown in colors. Solid (dashed) curves are convolutions with filament distribution $\mathcal{N}_\mathrm{f} \propto m_\lambda^{-2.2 (-2.5)}$. The integration is performed from MpL $=50~\Mspc$ to upper limits between $=100~\Mspc$ and $10000~\Mspc$ specified in the legend. The Chabrier IMF \citep{Chabrier05} shifted by a factor three in mass is shown in black. The results at masses lower than $10^{-2} ~\Ms$ are not shown since these small scale fluctuations are of rather high density and the barotropic eos representation is no longer physical.}
\label{fig_conv_hydro}
\end{figure}
Figure \ref{fig_conv_hydro} portrays the convolved CMF, as given by equation (\ref{eq_conv}), for a population of non-magnetized filaments. 
We take filaments of uniform length 10 pc and diameter 0.1 pc. 
The length of the filament does not have a strong impact on the results and the model is rather insensitive to this parameter. Since the maximal fragment size is limited to a few times the filament width, any filament longer than this value should exhibit similar fragmentation patterns. We assume a constant width for the filaments, as suggested by observations, but the model is easily scalable to varying filament widths. 
We consider filaments in global virial equilibrium \citep{Arzoumanian13}. 
As mentioned earlier, some observations suggest that the filament width is possibly increasing with the MpL \citep{Hill11}. As discussed in Appendix \ref{appen_R},
this does not affect significantly the results. 
Two distributions are considered for the FLMLF power-law exponent, $dN/dm_\lambda \propto m_\lambda^{-\alpha}$, namely 2.2 (solid) and 2.5 (dashed). 
As shown in the figure, the number of fragments per unit length increases and the peak mass decreases with increasing filament MpL, indicating that
the low-mass end of the convolved CMF is dominated by cores produced in massive filaments. 
We thus perform the integration from the filament of $50~\Mspc$ (equivalent to $\mathcal{M}_\mathrm{1D}=1$, note that filaments with higher density have higher temperature with the barotropic eos) to various upper limits between $100~\Mspc$ and $10000~\Mspc$ for each of the two distributions. 
These values bracket the most massive filaments reported by \citet{Arzoumanian13}. 
In the absence of magnetic field, the massive filaments fragment into numerous very small fragments. 
Whether filaments can get rid entirely of the magnetic field while accumulating such mass, however, is highly questionable. 
When limiting the calculations to filaments less dense than $100~\Mspc$, 
the peak mass is determined by the densest filaments and lies around $1~\Ms$. 
When denser filaments are included, the convolved CMF exhibits a slightly shallower slope.

Therefore, in the absence of magnetic field,
the peak of CMF decreases with increasing MpL and the slope of the final CMF is determined essentially by the filaments with the highest MpL, rather than by equation (\ref{eq_N_mod}).

\subsubsection{Magnetized case}
\begin{figure*}[]
\centering
%\plottwo{convolution_Ma8.pdf}{convolution_Ma5.pdf}
%\plottwo{convolution_Ma2.pdf}{convolution_Ma1.pdf}
\includegraphics[trim=3   53 3 12,clip,width=0.49\textwidth]{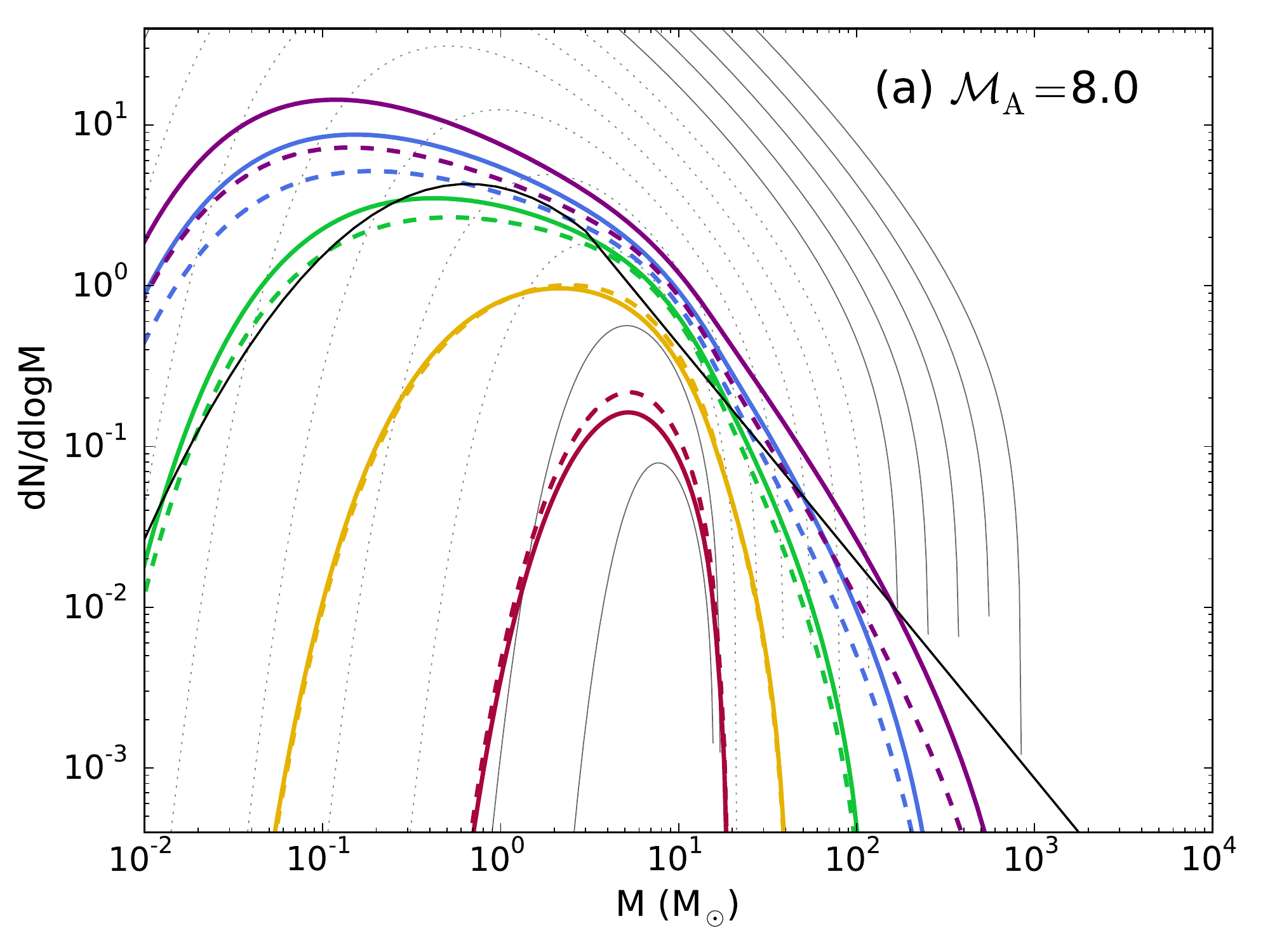}
\includegraphics[trim=3   53 3 12,clip,width=0.49\textwidth]{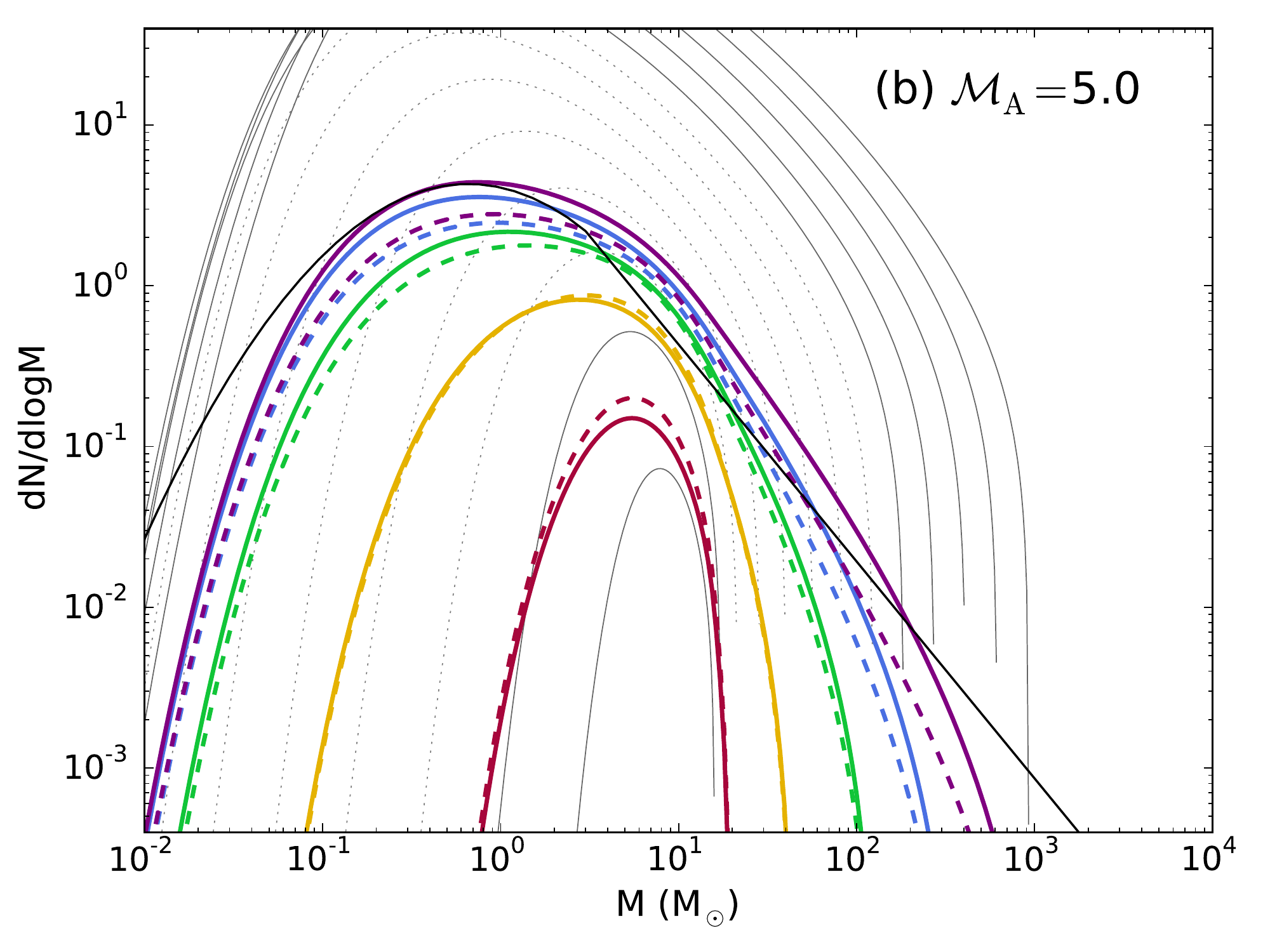}
\includegraphics[trim=3   8   3 12,clip,width=0.49\textwidth]{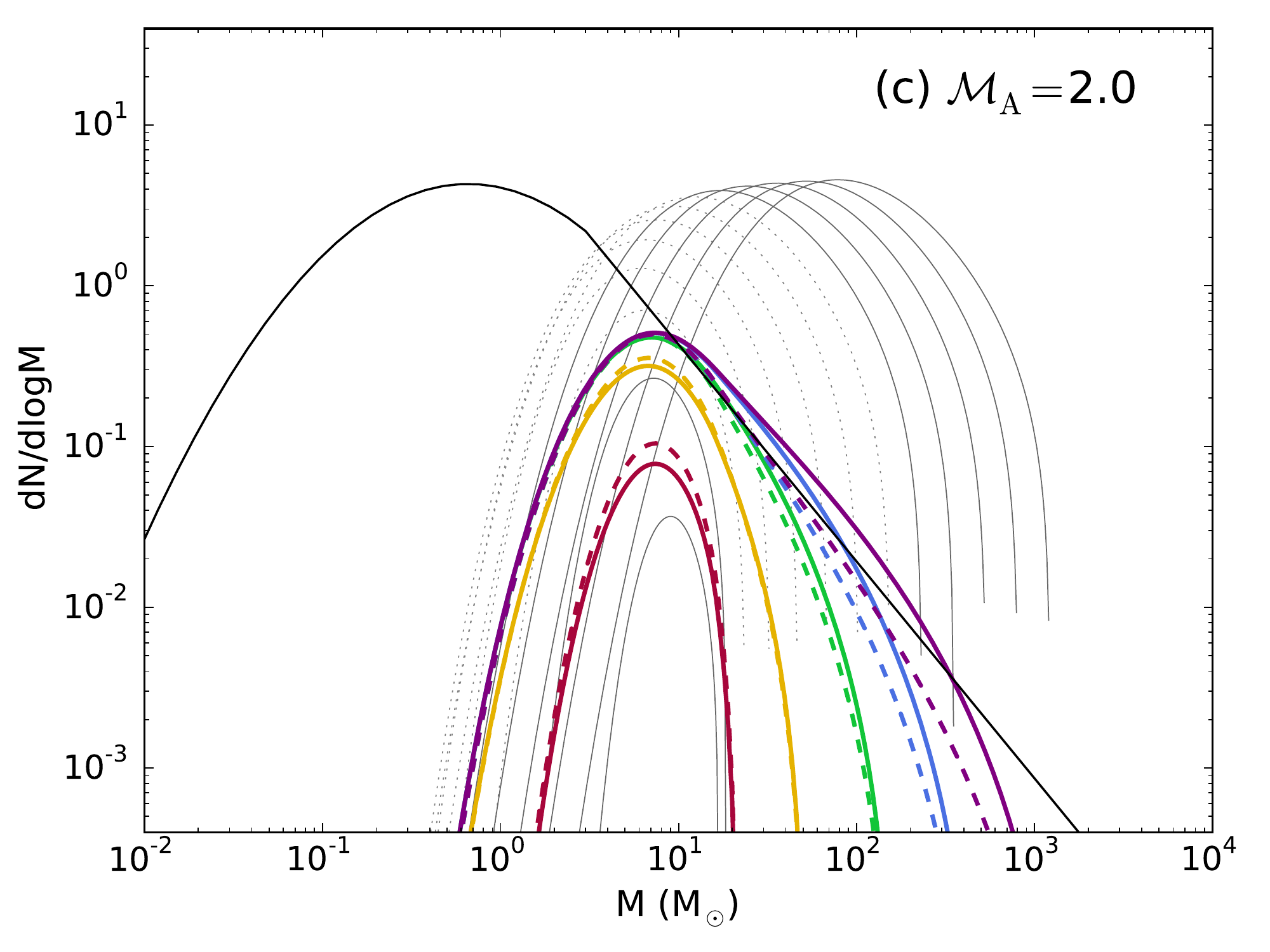}
\includegraphics[trim=5   8   4 12,clip,width=0.49\textwidth]{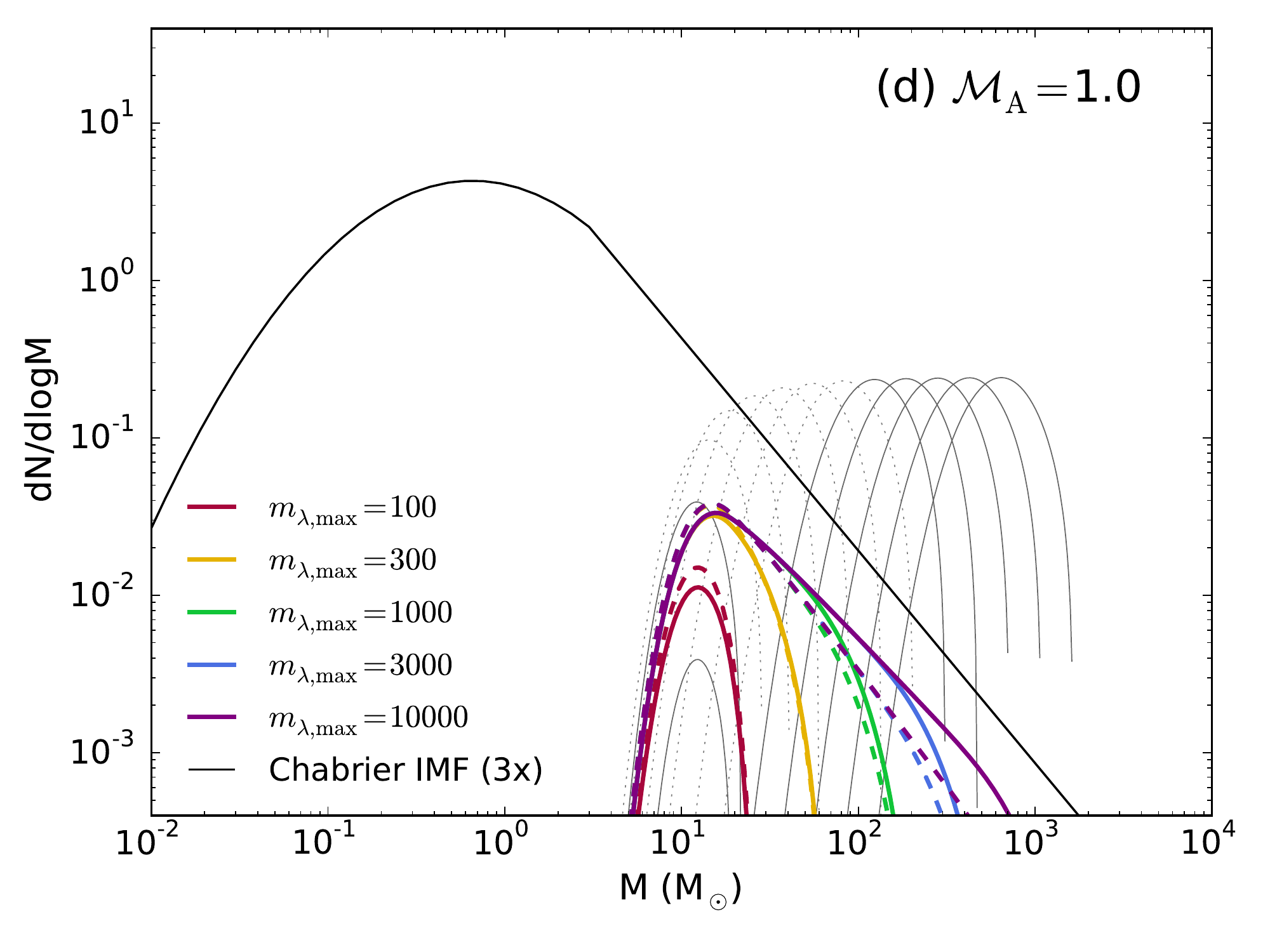}
\caption{Convolution for magnetized filament populations with $E_\mathrm{mag}/E_\mathrm{nt} = 0.015, 0.038, 0.2 \text{ and } 0.5$, corresponding to $\mathcal{M}_\alfv =$ 8 (a), 5 (b), 2 (c), and 1 (d), respectively. Same convolutions and color coding as that in Figure \ref{fig_conv_hydro}.}
\label{fig_conv_mag}
\end{figure*}
Same conditions as in Section \ref{st_conv_hydro} are considered, 
except that we include the magnetic field, in about energy equipartition with turbulent energy. 
Four values of magnetic over turbulent energy are considered, %namely $E_\mathrm{mag}/E_\mathrm{nt} = 0.015,~0.038,~0.2 \text{ and } 0.5$. 
corresponding to $\mathcal{M}_\alfv = 8, ~5, ~2 \text{, and }1$. %, respectively.
The results are displayed in  Figure \ref{fig_conv_mag}. 
Compared with the previous, pure hydro case, we note that the magnetic field allows the formation of more massive fragments, while decreasing the total number of fragments. 
A striking result is that only when filaments with high MpL are included, does the CMF recover a Salpeter-like slope at large masses. 
This implies that magnetized massive filaments, with magnetic field energy in rough equipartition with turbulent energy, are necessary to reproduce the observed universal CMF. 
As anticipated from the hydro case, for low level of magnetization,  
the low-mass end of the convolved CMF is dominated by filaments of high MpL, 
and thus the peak position varies with the upper mass limit of integration in equations.(\ref{eq_conv}). 
At $\mathcal{M}_\alfv \lesssim 5$, the impact of magnetic field becomes important enough 
for the peak mass to become basically independent of the integration over the filament population, 
and a reasonable agreement with the Chabrier IMF \citep{Chabrier05} is found. 
%As expected, the increasing level of magnetization lowers significantly the star formation efficiency. 

Therefore, our results suggest that in the case of prestellar core formation by filament fragmentation, 
the magnetic field plays a dominant role in the determination of the CMF. 
Meanwhile, the Alfv\'enic Mach number $\mathcal{M}_\alfv \sim 5$, 
slightly lower than energy equipartition with turbulence, 
can perhaps be naturally produced in turbulent MHD flows, and is worth further numerical investigation. 
Indeed it is compatible with the simulation results of \citet{Lee16a} (their Figure 11) of clustered environment, 
in which the magnetic energy is found to be around a few percent of the turbulent energy. 
A major test of the present theory would be to observe the distribution of bound prestellar cores in a non or weakly magnetized filament, where - all the other conditions been the same ! - we predict the number of cores to be larger and the CMF to peak at smaller masses than for magnetized filaments. 

We have shown results with $\gamma_\magb=0.5$ for magnetized filaments. 
This value is supported both by observations \citep{Crutcher10,Li15,LiP15} and simulations \citep{Kim01,Banerjee09} and thus appears to be a reasonable choice for studying the effect of magnetization on filament fragmentation. 
There is no analytical expression of the density PDF width for $\gamma_\magb \ne 0, 0.5$ and 1 and, for such cases, the mathematical (numerical) resolution becomes cumbersome. For $\gamma_\magb=0$, i.e. magnetic field independent of density, the density PDF is not modified by the magnetic field, and the Alfv\'en velocity scales as $v_\alfv^2 \propto \rho^{-1}$, which is very close to the turbulence scaling $v_\turb^2 \propto l^{2\eta_\turb} \propto \rho^{2\eta_\turb/(2\eta_\turb-2)}$ at large scales and dominated by the thermal energy at small scales. The magnetization of the filament, therefore, hardly modifies the CMF compared to the hydro case (for virialized filaments of same MpL). The case $\gamma_\magb=1$ ($B\propto \rho$) is very unlikely and not worth discussing. The variation of the CMF with $\gamma_\magb$, on the other hand, should be continuous, and we expect the resulting CMF for $0<\gamma_\magb<0.5$ to lie between these two extreme cases. Note, however, that the modification of the CMF by the magnetic field stems essentially from the narrowing of the density PDF, since the filament is globally virialized and the total amount of energy is unchanged and $\gamma_\magb$ has to be large enough to have an effect on the fragmentation. 
%The fragmentation behavior is sensitive not only to the amount of the supporting energy, but also to its nature.

\subsection{Uncertainties of this model}
Although the physical concept of our model is not particularly complicated and relies on the Hennebelle-Chabrier approach of gravo-turbulent fragmentation, the filamentary nature of the collapse introduces several 
uncertain parameters in the model. The filament fragmentation model itself is relatively straightforward and can be tested by counting cores inside filaments of same conditions, provided large enough statistics is available. The largest uncertainties in this study appear when introducing a population of filaments to calculate the final CMF. Even though the parameters have been chosen to be consistent with observations, they can of course be better determined with more observational constraints. The filament length, but most importantly width and MpL, which have a direct impact on the fragmentation, should be relatively easy to determine observationally. 
%The length is practically important for the integration but does not affect much the fragmentation.} 

The most uncertain parameter is the magnetic field strength: whether it depends on the MpL and whether there is a spread around some typical value is presently unknown. Therefore, we suggest using the present model to 
better constrain the level of magnetization  (or any other equivalent source of pressue support)  within filaments from the confrontation between the predicted theoretical CMF and the observed one. 
 Detailed modeling of the full MHD dynamics may reveal effects that are neglected in the current study. 

%________________________________________________________________________________

\section{Conclusions}
In this paper, we have derived a theory of mass fragmentation inside filamentary molecular clouds using the Hennebelle-Chabrier formalism adapted to filamentary geometry. 
We consider thermally supercritical filaments with a wide range of MpL, in cylindrical virial equilibrium, subject to various levels of turbulence or magnetic field strength. 
This introduces a scale dependence in the energy cascade and thus in the (virial) collapse condition.
The CMF derived for individual filaments is then convolved with different distributions of filament populations to obtain the final system CMF.
The main conclusions of our calculations are as follows:

\begin{itemize}

\item Purely hydrodynamic filaments with high MpL supported by turbulence tend to fragment into too small pieces, meaning that the CMF is bottom-heavy in high MpL filaments. This is at odds with present observations. To hamper such small scale fragmentation, the massive filaments must be moderately magnetized  (or have some additional pressure support from e.g.\ cosmic rays or proto-stellar heating, although so far observations suggest that none of these effects seem to have the right intensities 
) , i.e. with Alfv\'enic Mach number of a few, i.e. mass-to-flux ratio a few times the critical value. 

\item Filaments with high MpL are needed to produce massive cores. Without the presence of high MpL ($>1000 ~\Mspc$) filaments, the convolved CMF slope is steeper than the Salpeter value.

\item The filamentary geometry naturally introduces a change in behavior at different scales. While at small scales, the fragmentation is essentially a 3D (spherical) process, the process is almost one-dimensional at large scales, when reaching the turbulence-dominated 1D regime. In the present model, this change in the geometry naturally leads to two modes of star formation. Indeed, since cores form mostly at scales smaller than the filament width, the geometry imposes an upper mass cutoff for their formation. Small filaments, on one hand, directly fragment into small (spherical) cores, as described in the standard HC formalism. Large filaments, on the other hand, follow a two-level hierarchical fragmentation process, as suggested by observations \citep{Kainulainen13,Hacar13,Takahashi13,Pineda13,Teixeira16,Kainulainen17}. Indeed, within large filaments, instability at the largest scales (the first-crossing) yields a mass distribution of clumps, or group of cores, which displays a powerlaw at high masses whose exponent strongly depends on the filament MpL and   extra pressure support (most likely magnetic pressure) . This distribution becomes narrower in the presence of magnetic field. These self-gravitating clumps then undergo sub-fragmentation that yields the CMF (last-crossing). 
The present model thus points to a 2-mode process for star formation.

\item Four levels of magnetization are discussed: $\mathcal{M}_\alfv = 8, ~5, ~2$ and $1$. When convolved with the filament distribution, $dN/dm_\lambda \propto m_\lambda^{-2.2\sim2.5}$, the final system CMF is reasonably similar to the observed one. Strikingly enough, already for modest levels of magnetization, $\mathcal{M}_\alfv \sim 5$, the CMF peak position becomes independent of the integration over the filament population. 

\item Convolving the core mass function within a given filament with a filament population leads ultimately to the final CMF in the cloud. Interestingly enough, our results suggest a minimum level of 
 pressure support of filaments $c_{\rm eff}^2 = P_{\rm total}/\rho \gtrsim v_\turb^2 / 5$ (e.g. $\mathcal{M}_\alfv \lesssim 5$, if magnetic field provides this support) 
%magnetization of filaments $\mathcal{M}_\alfv \lesssim 5$ 
in order to reproduce a CMF similar to the observed one regardless of the filament population. We find that the results are robust to variations of filament MpL properties (slope and MpL cutoff), and best fit the Chabrier IMF for $\mathcal{M}_\alfv \sim 5$. Our theory, therefore, provides a viable explanation for the lack of variation of the IMF, under the assumptions that the properties of
the filaments are sufficiently universal and  that there is a direct mapping from the CMF to the IMF. Indeed this model does not address the transition from the CMF to the IMF, which might involve further gravitational fragmentation.

\item There is a lack of low-mass stars in highly 
 magnetized filament populations that meet the above criterion. 
%magnetized filament populations with $\mathcal{M}_\alfv < 5$. 
We discuss several possibilities to solve this problem. 
\begin{itemize}
\item Whereas most stars appear to form within filaments \citep[75\% as suggested by][]{Konyves15}, others (mostly low-mass ones, produced by small-scale density fluctuations) might form through the standard (spherical) Hennebelle-Chabrier fragmentation process. Therefore, our model supports the idea of a two-mode star formation process. This is also consistent with the finding that no low mass cores are found in the very massive filament W43 (Motte et al private communication).
\item A lognormal density PDF is used inside the filament throughout the calculations, whereas powerlaw tails are often observed to develop under the action of self-gravity. This will induce further gravitational fragmentation into smaller cores  compared to the present model. High density fluctuations, leading to low-mass objects, correspond to small scales, for which thermal support dominates. A powerlaw PDF $\propto \rho^{-\alpha}$ yields a mass spectrum $\mathcal{N} \propto M^{-2(2-\alpha)}$ from equation (\ref{eq_MF}) at low masses. This means no low mass turn off for values of $\alpha <2$, the usually observed value. 
Note, however, that the powerlaw tail, due to gravity, starts to develop only after some cores become self-gravitating 
%and modifies only the density within the collapsing core, 
and thus should not drastically modify the CMF, even though we can not say exactly
to which extent it affects the fragmentation process.

%\item There could be a distribution of Alfv\'enic Mach number which is invariant within a sufficiently large sample of filaments and thus yielding a universal CMF.
\end{itemize}

\end{itemize}

\begin{acknowledgements}
   This research has received funding from the European Research Council under
   the European Community's Seventh Framework Programme (FP7/2007-2013 Grant
   Agreement no. 306483). The authors thank helpful and stimulating discussions with P. Andr\'e, V. K\"onyves, and D. Arzoumanian. The authors also thank the anonymous referees that have helped substantially improve the manuscript. 
\end{acknowledgements}

\appendix

%________________________________________________________________________________
\section{The smoothing of the geometry transition}\label{appen_n}
Throughout this study, the value $n=2$ is used for equation (\ref{eq_r_of_l}). 
This smoothing allows a transition from the 3D spherical regime to the 1D cylindrical regime by using a ellipsoidal description throughout the scales. 
The analytical motivation is to express the fragmentation with one single smooth function. 

At the same time, there are several physical motivations. 
Firstly, at scales much smaller than the filament width, 
the fragmentation should be insensitive to the geometry and thus purely spherical, 
while the cores start to feel the shape of the filament when they approach scales comparable to the filament width. 
This should be a smooth transition to the linear regime where mass increases with the size only along the filament axis. 
Secondly, we use an ideal cylinder of uniform density in this study, that certainly deviates somehow from the reality. From observations, 
thermally supercritical filaments usually have a central density plateau and a wing following a powerlaw relation \citep[$\rho \propto r^{-2}$ in][]{Arzoumanian11}. 
This implies that when increasing the size in the filament radial direction, 
the increase in mass slows down due to decrease in density. 
This also requires a smoothing description for the formalism. 
Here we demonstrate that using the simple smoothed representation in equation (\ref{eq_r_of_l}) is comparable to considering a filament with radial density profile, 
and that $n = 2$ is a reasonable number to use. 

Let us consider a filament of radius $R$ with radial density profile
\begin{equation}
\rho(r) = {\rho_0 \over 1+\left({r \over r_0}\right)^2},
\end{equation}
where $r_0$ is the radius of the inner plateau and $r$ the distance to the axis. 
A sphere of radius $l$ centered at the filament axis has mass
\begin{eqnarray}
\label{eq_r_e}
M(l) &=& \int\limits_0^{\min(l,R)} \rho 2\pi r 2 \sqrt{l^2-r^2} dr \\
&=& 4\pi \rho_0 l r_0^2  \left\{ \sqrt{1-\left({R\over l}\right)^2} -1 - \sqrt{{\left({r_0 \over l}\right)}^2+1} \left[ \tanh^{-1}\sqrt{1-\left(R\over l\right)^2 \over \left({r_0 \over l}\right)^2+1} -  \tanh^{-1}\sqrt{1 \over \left({r_0 \over l}\right)^2+1} \right] \right\}   \nonumber \\
&=&{4\over 3} \pi  \rho_0 l r_\mathrm{e}^2, \nonumber
\end{eqnarray}
where $r_e$ is the equivalent semi-minor axis of an ellipsoid of uniform density. 
We plot in Figure \ref{fig_smoothening} $r$ defined by equation (\ref{eq_r_of_l}) for several $n$ values, 
and $r_e$ defined by equation (\ref{eq_r_e}) for several ratios between $r_0$ and $R$. 
The central density $\rho_0$ is used for estimating the equivalent semi-minor axis. 
The reason that this value is used instead of the average density is that we calculate only mass contained in spheres centered on the filament axis, 
while in reality there exists a distribution of positions that results in lowered masses that are not taken into account. 
These two effects roughly cancel out, 
thus we choose this simple expression without going too much into the details. 

\begin{figure}[]
\centering
\includegraphics[trim=0 0 0 0,clip,width=0.5\textwidth]{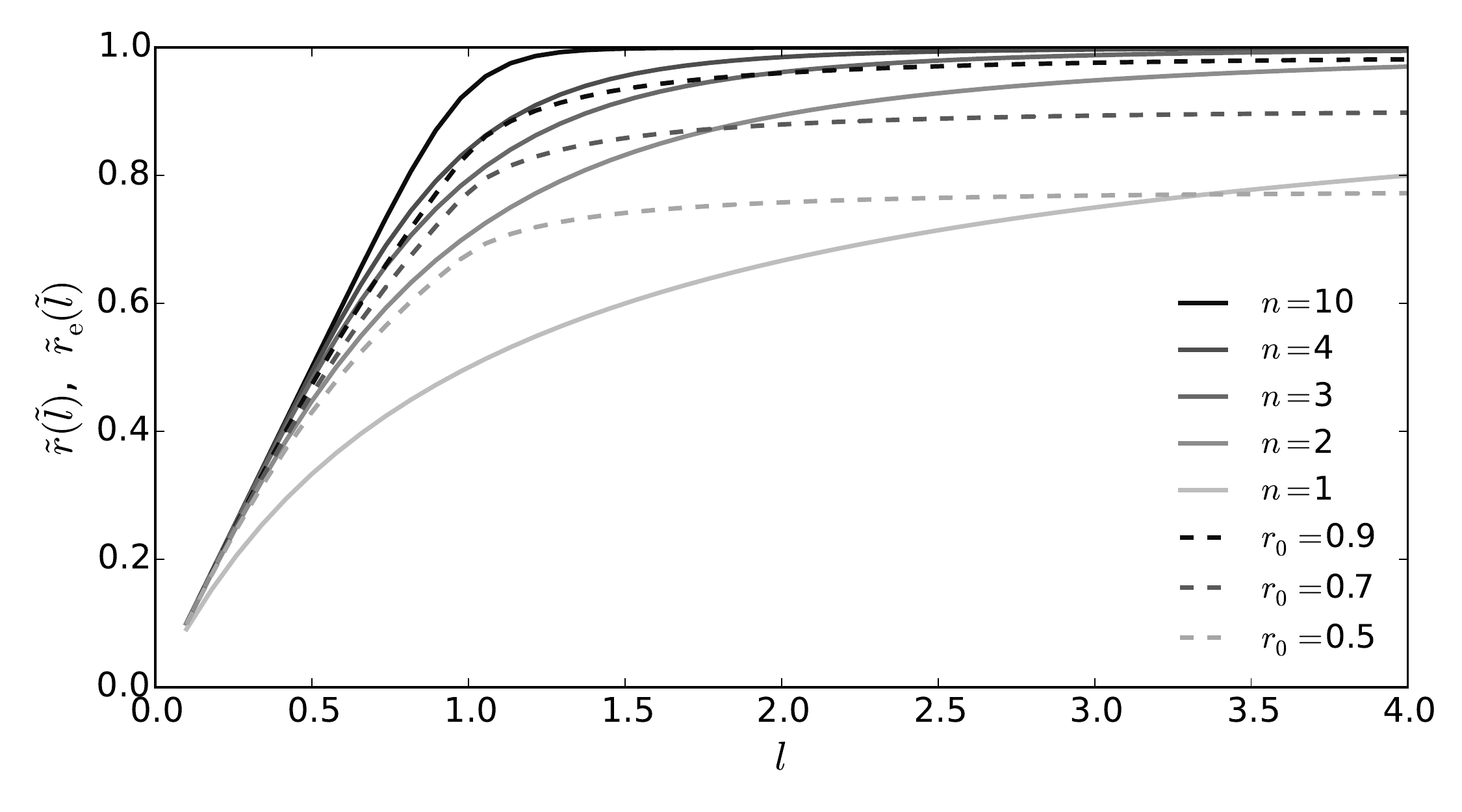}
\caption{The clump radius used in our model $\widetilde{r}(\widetilde{l})$ in equation (\ref{eq_r_of_l}) is plotted with the smoothing parameter $n = 1,~2,~3,~4$ and $10$ in solid curves from bottom to top. The equivalent radius $\widetilde{r}_\mathrm{e}(\widetilde{l})$ inferred from filament with a radial density profile in equation (\ref{eq_r_e}) is plotted for $r_0/R = 0.5,~0.7,$ and $0.9$ in dashed curves form bottom to top.}
\label{fig_smoothening}
\end{figure}

\begin{figure}[]
\centering
\includegraphics[trim=0 0 0 0,clip,width=0.5\textwidth]{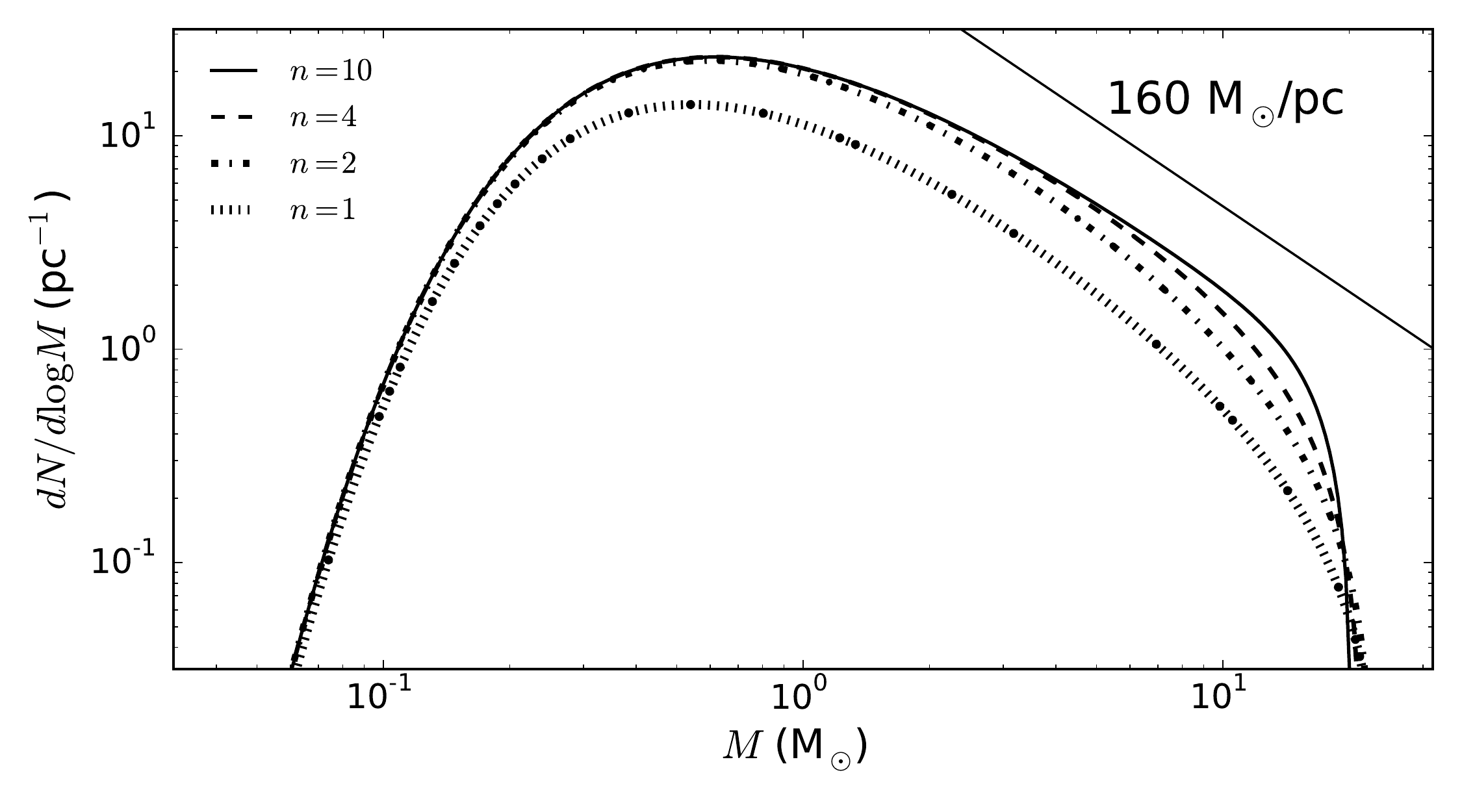}
\caption{CMF of $160~\Mspc$ non-magnetized filament with smoothing parameter $n=1,~2,~4,$ and $10$. The smoothing with $n=1$ has a strong effect on the form of the CMF, while with $n \gtrsim 2$, the variation remains limited.}
\label{fig_ns}
\end{figure}

We are more concerned about low $\widetilde{l}$ values at which self-gravitating cores from. 
With $n \gtrsim 2$, equation (\ref{eq_r_of_l}) reasonably approximates the behavior of equation (\ref{eq_r_e}) for $r_0 /R \gtrsim 0.5$. 
This implies that with our simplified formalism, we are describing the central part of the filament including a region slightly larger than the central density plateau. 
We also show in Figure \ref{fig_ns} the CMF of a $160~\Mspc$ non-magnetized filament with different values of $n$. 
As long as $n \gtrsim 2$, the choice of its value does not have a strong impact on the model results. 
It is important to note that equation (\ref{eq_r_of_l}) is a mathematical choice that we made to illustrate the transition from spheres to cylinders. 
Other possible descriptions can be used as long as physically motivated. 
Nonetheless, this simple formula captures reasonably well the effects of the geometrical transition.

\section{Powerlaw envelope of the filament}\label{appen_p}
In this work we considered filaments of constant density, while observed filaments often exhibit a central plateau plus a powerlaw envelope. Here, we determine the correction of the unstable critical mass when such an envelope is taken into account. Let us consider the density profile in eqn.(\ref{eq_plummer})
where the exponent $p$ is observationally reported to vary between 1.5 and 2.5 \citep[e.g.][]{Arzoumanian11}, and usually $p \lesssim 2$. Now, instead of considering a prolate ellipsoid of semi-major axis $l$ (described by eqn.(\ref{eq_r_of_l})), the clump is now a sphere of radius $l$ and encompasses mass from the filament core (of radius $r_0$) and the powerlaw envelope (For simplicity the clump is always centered on the filament axis). The gravitational energy in eqn.(\ref{egrav}) should be re-written:
\begin{equation}\label{egrav2}
-E_\grav(M_l) = {3 \over 5} {GM_l^2 \over l} w_\grav(1) \varpi_\grav(l).
\end{equation}
In this case, the geometrical factor $w_\grav(1)=1$, while an extra factor $\varpi_\grav(l)$ is introduced to describe the level of mass concentration, compared to a uniform density profile. To understand the effect of the filament powerlaw envelope, we are interested in cases where $l \gtrsim R$, while $R \gtrapprox r_0$, as described in Appendix \ref{appen_n}. 
For simplicity, we assume that the supporting energy is exactly the same as in the uniform density model, since the turbulence evaluated at $l$ and magnetic Alfv\'{e}n velocity at the central density are reasonable estimates. For a fixed mass, the gravitational energy is decreased due to the absence of the ellipsoidal geometry as in the uniform density case, but at the same time it is increased as a result of the density profile. As shown in Fig. \ref{fig_w_g}, $w_\grav$ goes to $\sim 3$ at $\eta \sim 10$, which is a reasonable upper limit of filament envelope $r_{\rm e}/r_0$. 
We do not compute the exact value of $\varpi_\grav(l)$, but is is easy to see that it approaches 1 at small $l$ and is slightly larger than 1 at larger $l$ (the value is 5/3 for a sphere with pure $r^{-2}$ density profile, which could be regarded as a reference upper limit since in this case there is a central plateau and the powerlaw concerns only the cylindrical radial direction), although this effect is more pronounced for larger $p$ values. 
When applying the virial equilibrium condition, the critical mass at a given scale $l$ differs by the factor %$$
\begin{equation}
{M_{\rm profile}(l) \over M_{\rm uniform}(l)} = {w_\grav(l) \over \varpi_\grav(l)}.
\end{equation}
The effects more or less cancel out and give a ratio $\lesssim 3$. 

On the other hand, the corresponding critical density (at the filament axis) should be about the same, since part of the mass comes from the envelope. Altogether we can see that, when considering the powerlaw envelope of the filament, the quantities entering eqn.(\ref{eq_MF}) are not drastically affected, except for large $l$ for which the mass spectrum is shifted to higher mass by a small factor $\lesssim 3$. Finally, the right panel of Fig. \ref{fig_cases} shows that clumps of sizes larger than the filament width are extremely unstable to fragmentation. It is thus clear that considering a realistic filament with a powerlaw density envelope introduces some correction to the group mass function derived in this work, by hardly modifies the final CMF.

\section{The gravitational potential energy of an ellipsoid}\label{appen_w}
Following the calculations of \citet{Neutsch79}, 
an ellipsoid with semi-axes $R, R, R\eta$ with uniform density has gravitational potential energy
\begin{equation}
E_\grav = {3\over 5} {G M^2 \over \eta R} w_\grav(\eta),
\end{equation}
where $G$ is the gravitational constant and $M$ the total mass. 
The function $w_\grav(\eta)$ is a geometrical factor
\begin{equation}
w_\grav(\eta) = \left \{ \begin{array} {cll}
{\eta\cos^{-1}(\eta) \over \sqrt{1-\eta^2}} &={\eta\sin^{-1}(\sqrt{1-\eta^2}) \over \sqrt{1-\eta^2}}& ,~ \eta < 1 \text{ oblate}\\
{\eta\log{(\sqrt{\eta^2-1}+\eta)} \over \sqrt{\eta^2 -1}} &= {\eta\sinh^{-1}(\sqrt{\eta^2 -1}) \over \sqrt{\eta^2 -1}}& ,~\eta>1 \text{ prolate}.
\end{array} \right.
\end{equation}
We express $w_\grav$ piece-wisely, while it is indeed a smooth function of $\eta$. 
The ellipsoid is oblate when $\eta <1$, and  prolate when $\eta>1$. 
With our parametrical setup that always $l>r$, we are using the $\eta > 1$ regime of this function. 
In Figure \ref{fig_w_g}, we plot $w_\grav$ as well as its derivative to illustrate the dependence on $\eta$.

\begin{figure}[]
\centering
\includegraphics[trim=0 0 0 0,clip,width=0.5\textwidth]{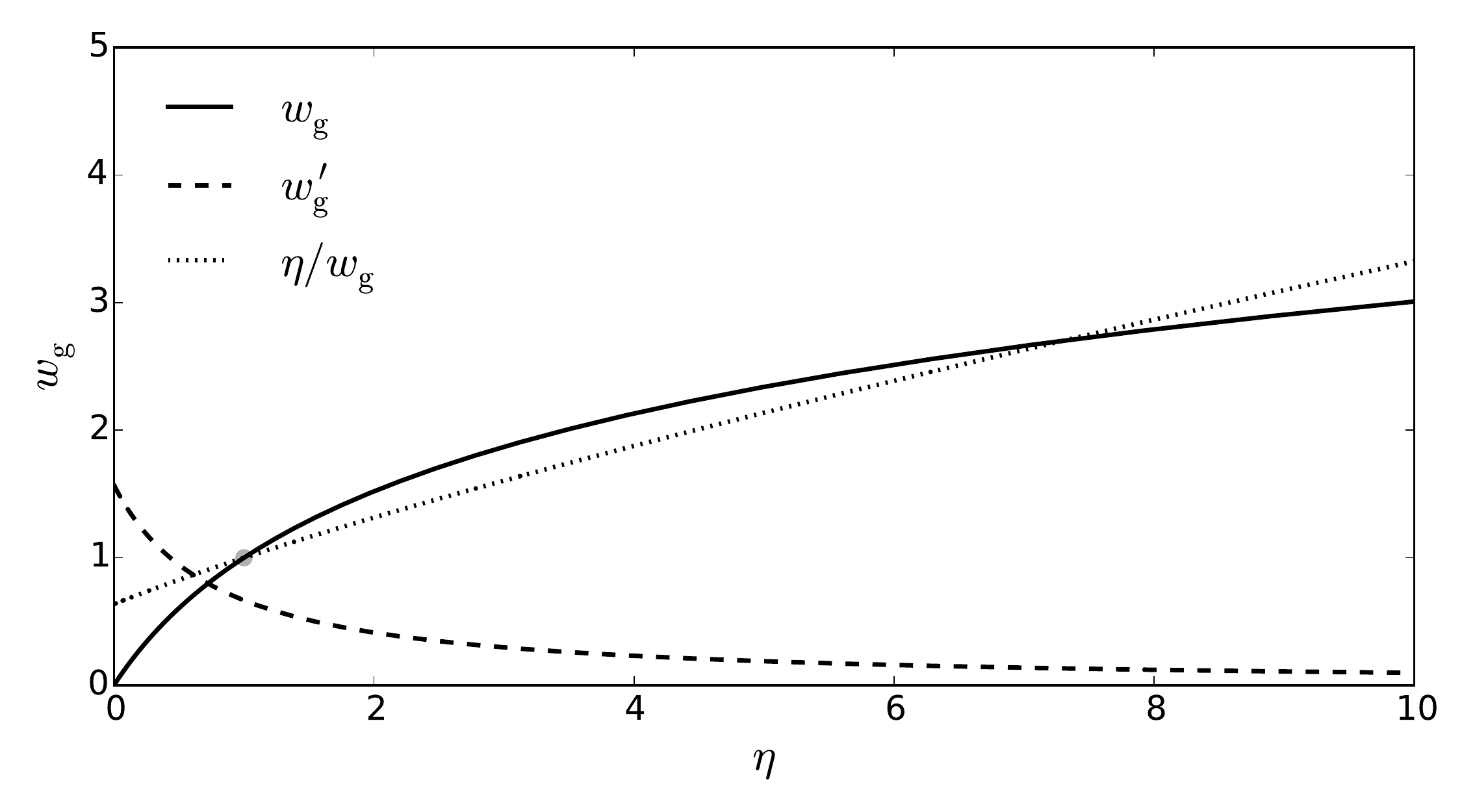}
\caption{The geometrical factor $w_\grav(\eta)$ (solid curve) and its derivative (dashed curve). The value of $\eta/w_\grav$ is also plotted (dotted) for better understanding of the equation behaviors. The circle shows where the function passes $w_\grav(1)=1$.}
\label{fig_w_g}
\end{figure}

\section{Mass function of groups}\label{appen_g}
The mass function of groups, that results from longitudinal fragmentation of filaments, 
is a less discussed subject. 
As mentioned in Section \ref{st_Hierarchical}, we are entering the one-dimensional regime at scales $\gtrsim R$. 
The fragmentation is constrained along the filament longitudinal direction and the result represents the  groups of cores. 
In this regime, the two terms in the right-hand-side of equation (\ref{eq_MF}) becomes comparably important. 
We thus show them separately to illustrate their behaviors. 

\begin{figure}[]
\centering
\includegraphics[trim=0 0 0 0,clip,width=\textwidth]{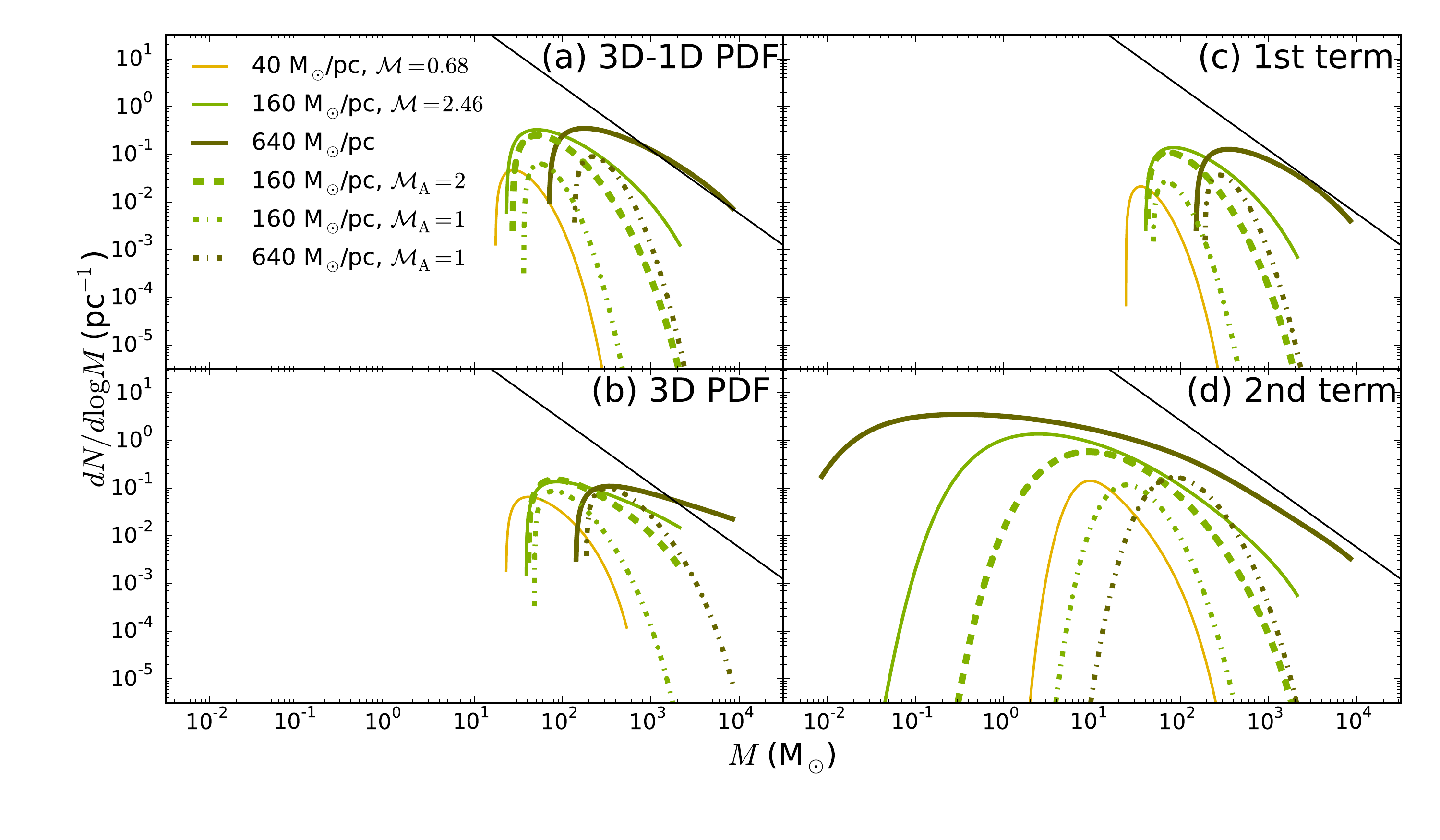}
\caption{(a) Mass functions of groups of cores with the canonical model, same as that in Figure \ref{fig_cases} with same color codings. (b) Mass functions of groups of cores without dimensional correction of density PDF width at the 1D regime. (c) First term in equation (\ref{eq_MF}). (d) Second term in equation (\ref{eq_MF}).  The black line shows the Salpeter slope $M^{-1.33}$.}
\label{fig_groups}
\end{figure}

In the upper left panel of Figure \ref{fig_groups}, we show the mass function of groups, 
same as that  in the lower left panel of Figure \ref{fig_cases}. 
The two terms in the right-hand-side of equation (\ref{eq_MF}) are shown in the upper and lower right panels. 
The first term is slightly dominant, 
but they are indeed of comparable amplitude. 
At smaller scales, the negative first term becomes largely dominant and determines the lower mass cutoff of groups. 

In the lower left panel we show the mass function of groups without the density PDF correction from 3D to 1D in equation (\ref{eq_dPDF}). 
Without the narrowing of the PDF due to restriction in the radial direction when going towards larger scales, 
the slope of the mass function becomes shallower, that is reasonably expected. 
Having a realistic description of the density PDF is, therefore, essential, 
and we need constrains from theories as well as observations.

\section{A probable scenario for filament formation}\label{appen_y}
No theoretical discussion has yet been done on the FMLF, that has an almost Salpeter slope of 
$dN/d m_\lambda \propto m_\lambda^{-2.2}$ for supercritical filaments (\citet{Andre14}; Arzoumanian et al. in prep.). 
Here we propose a scenario from which we infer this relation from anisotropic collapse of self-gravitating clumps.
The column density PDF of the Aquila cloud in Figure 15 of \citet{Konyves15} has a high-mass end slope very similar to that of the FMLF. 
Since this region is dominated by filamentary structures, the PDF is also dominated by the filament.
The similarity between slope values therefore implies that the filament length is nearly independent of its MpL, 
and thus the sum of filament lengths in a MpL bin should be proportional to the number of filaments. 

Firstly, we consider self-gravitating clumps following a mass distribution
\begin{equation}
\mathcal{N}_\mathrm{clump}(M) = \frac{dN_\mathrm{clump}}{dM_\mathrm{clump}} \propto M_\mathrm{clump}^{-(2+x)},
\label{eq_Nclump}
\end{equation}
where $x$ has typical value between 0.1 and 0.5 \citep{HC13}, around the Salpeter value 0.35. 
The idea is that these clumps are formed through gravo-turbulent processes. 
Secondly, the self-gravitating clumps collapse anisotropically in radial direction to form 0.1 pc wide filaments, 
while the longitudinal direction should also undergo some contraction. 
The number of filaments is proportional to the total filament length formed from one clump, 
that might have a slightly sub-linear dependence on the initial clump size
\begin{equation}
N_\mathrm{f} \propto R_\mathrm{clump}^{\zeta},
\label{eq_sublinear}
\end{equation}
where $ \zeta \lesssim 1$.
The self-gravitating clumps follow the mass-size relation 
$M_\mathrm{clump} \propto R_\mathrm{clump}^{1+2\eta_\mathrm{t}}$ in the supersonic regime before the collapse \citep{HC08}.
The MpL is thus
\begin{equation}
m_\lambda \propto \frac{M_\mathrm{clump}}{N_\mathrm{f}} \propto R_\mathrm{clump}^{1+2\eta_\mathrm{t}-\zeta} \propto M_\mathrm{clump}^{1+2\eta_\mathrm{t}-\zeta \over 1+2\eta_\mathrm{t}}.
\label{eq_mlambda}
\end{equation}
Equations (\ref{eq_Nclump}, \ref{eq_sublinear}, and \ref{eq_mlambda}), together, give
\begin{equation}
\mathcal{N}_\mathrm{f} ~=~ {dN_\mathrm{f} \over d m_\lambda} \\
~ \propto~  \frac{dN_\mathrm{f}}{dN_\mathrm{clump}}  \frac{dN_\mathrm{clump}}{dM_\mathrm{clump}} \frac{dM_\mathrm{clump}}{dm_\lambda} \nonumber\\
~ \propto ~M_\mathrm{clump}^{{2\zeta\over 1+2\eta_\mathrm{t}} -(2+x)} \nonumber\\
~ \propto~ m_\lambda^{-\left(2 + {x(1+2\eta_\mathrm{t}) \over 1+2\eta_\mathrm{t}-\zeta}\right)} 
~\propto~ m_\lambda^{-(2+y)}. \nonumber
\end{equation}

Given the canonical value of $\eta_\mathrm{t} = 0.45$, we plot in Figure \ref{fig_FMLF_y} the FMLF slope exponent $y$ in the range  $0.3 < \zeta <1$ for several values of $x$. 
Theoretical model \citep{HC13} predicts the high-mass-end slope to be shallower when the magnetic field is stronger, or when the cloud Mach number is small while the Mach number at Jeans length is large (their equation (25)), 
corresponding generally to conditions with lower density than the normalization of Larson's relation. 
The observed value of $0.2$ for $y$ corresponds to small $x$ values, 
and probably gives a hint on the filament formation environment. 
We note that this is a simplified proposition of filament formation, 
and how clumps collapse/fragment into filaments needs to be further investigated in details. 
\begin{figure}[]
\centering
\includegraphics[trim=0 0 0 0,clip,width=0.5\textwidth]{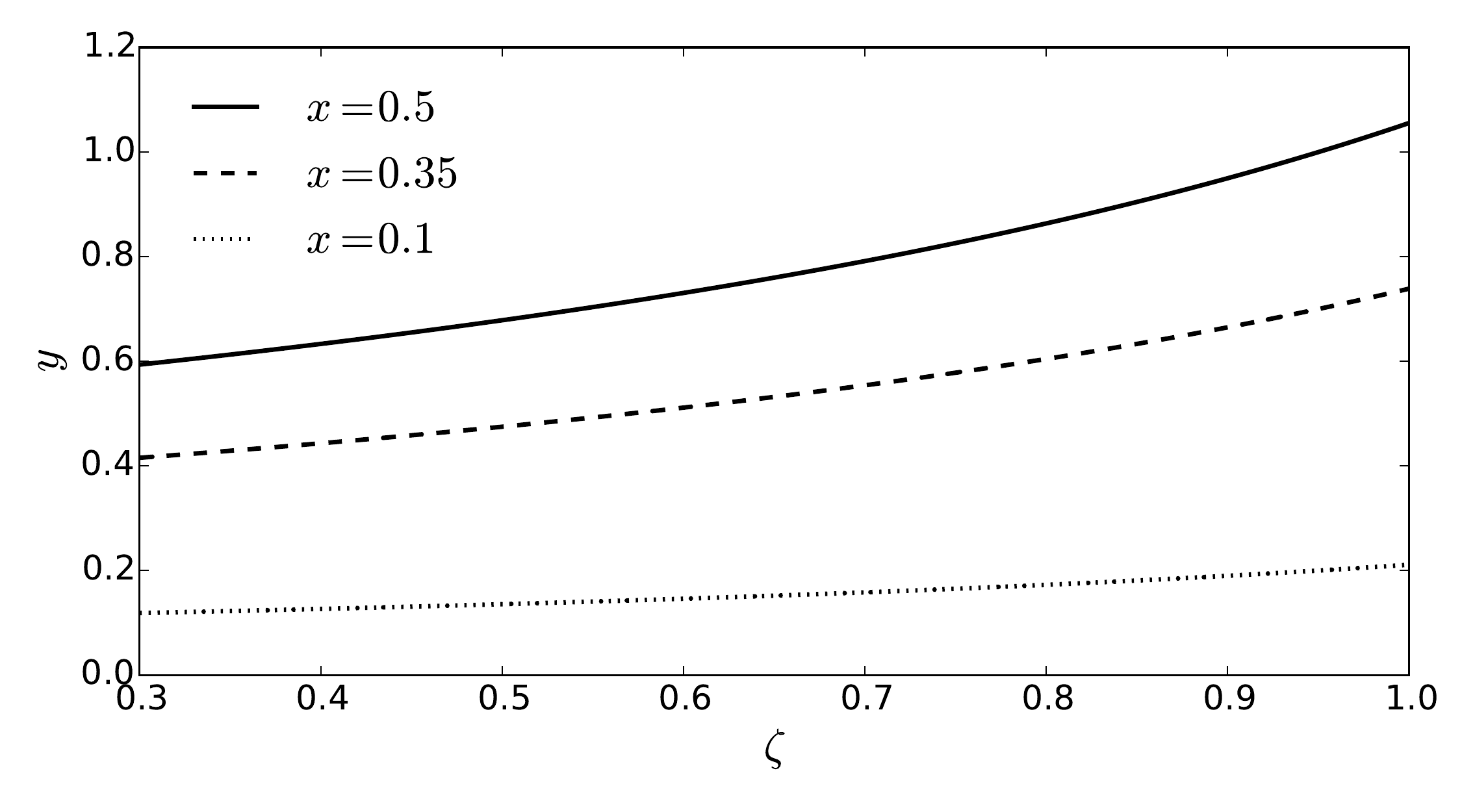}
\caption{Slope exponent of the FMLF inferred from self-gravitating clumps with $dN_\mathrm{f} / d m_\lambda \propto m_\lambda^{-(2+y)}$. The observational Salpeter value of $x = 0.35$ (dashed) as well as the theoretically derived range bracket 0.1 (dotted) and 0.5 (solid) are shown. The $x$ value increases from the bottom curve to the top curve.}
\label{fig_FMLF_y}
\end{figure}

\section{Non-uniform filament width}\label{appen_R}
\begin{figure}[]
\centering
\plottwo{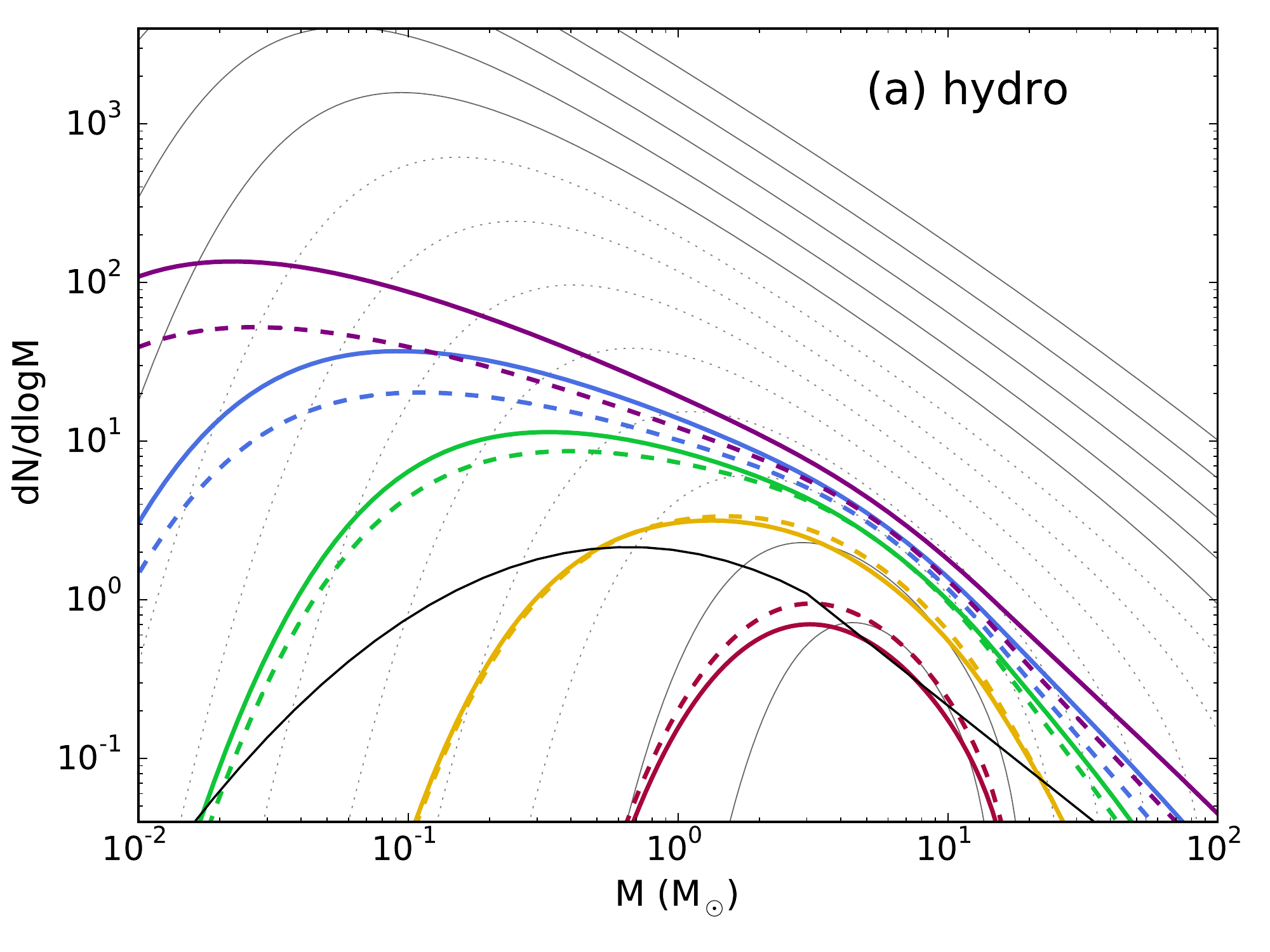}{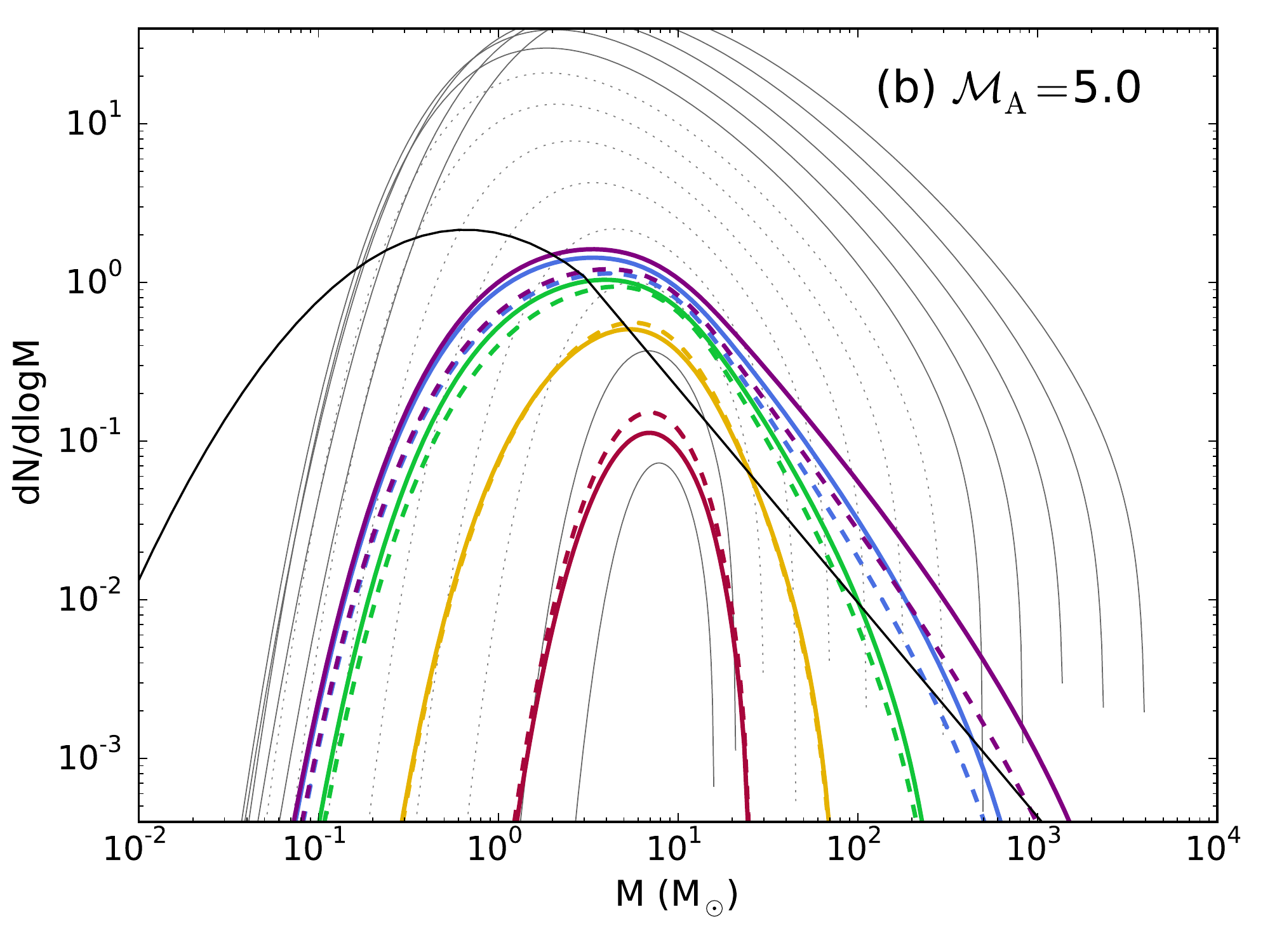}
\caption{System CMF of different MpL distributions with non-magnetized filaments (a) and magnetized filaments at $\mathcal{M}_\alfv=5$ (b). Color codings are same as that in Figure \ref{fig_conv_hydro}. The filament diameter varies linearly with the MpL from $0.1 ~\pc$ at $50 ~\Mspc$ to $0.5 ~\pc$ at $10000 ~\Mspc$. The purely hydro case has no significant difference from the distribution with uniform $0.1 ~\pc$, while the CMF slope of the magnetized case becomes slightly shallower when the filament width increases with increasing MpL.}
\label{fig_conv_w}
\end{figure}
As suggested by \citet{Hill11} from observations of Vela C, 
filaments with higher MpL are possibly wider. 
We show in Figure \ref{fig_conv_w} the convolved CMF for a population of non-magnetized filaments same as in section \ref{st_conv}, 
except that the filament width increases from 0.1 pc to 0.5 pc from MpL $50~\Mspc$ to $10000~\Mspc$. 
Same convolutions are shown for the hydro and $\mathcal{M}_\alfv=5$ cases. 
The widening of massive filaments increases their corresponding characteristic mass and thus shifts the peak towards slightly larger mass. 
But this effect is very small and is almost negligible in the non-magnetized case. 
When magnetized, the increase of width with MpL shallows slightly the convolved slope of the system CMF and shifts the peak towards larger mass. 
The effect is more significant when filaments with higher MpL are considered. 

\bibliographystyle{aasjournal}
\bibliography{biblio_filament}

\end{document}